\documentclass[]{emulateapj} 
\usepackage{placeins}
\usepackage{graphicx}
\usepackage[normalem]{ulem}
\usepackage{empheq}
\usepackage{tablefootnote}
\usepackage{amssymb, amsmath}
\usepackage{mathtools}
\usepackage{multirow}
\usepackage{xfrac}

\newcommand{\Ms}{M_{\odot}}
\newcommand{\npe}{$n$-$p$-$e$}

\begin{document}

\title{Finite-temperature extension for cold neutron star equations of state}
\author{Carolyn A. Raithel, Feryal \"Ozel, \& Dimitrios Psaltis}

\begin{abstract}
Observations of isolated neutron stars place constraints on the equation of state (EOS) of cold, neutron-rich matter, while nuclear physics experiments probe the EOS of hot, symmetric matter. Many dynamical phenomena, such as core-collapse supernovae, the formation and cooling of proto-neutron stars, and neutron star mergers, lie between these two regimes and depend on the EOS at finite temperatures for matter with varying proton fractions. In this paper, we introduce a new framework to accurately calculate the thermal pressure of neutron-proton-electron matter at arbitrary density, temperature, and proton fraction. This framework can be expressed using a set of five physically-motivated parameters that span a narrow range of values for realistic EOS and are able to capture the leading-order effects of degenerate matter on the thermal pressure.
We base two of these parameters on a new approximation of the Dirac effective mass, with which we reproduce the thermal pressure to within $\lesssim30\%$ for a variety of realistic EOS at densities of interest. Three additional parameters, based on the behavior of the symmetry energy near the nuclear saturation density, allow for the extrapolation of any cold EOS in $\beta$-equilibrium to arbitrary proton fractions.  Our model thus allows a user to extend any cold nucleonic EOS, including piecewise-polytropes, to arbitrary temperature and proton fraction, for use in calculations and numerical simulations of astrophysical phenomena. We find that our formalism is able to reproduce realistic finite-temperature EOS with errors of $\lesssim20\%$ and offers a $1-3$ orders-of-magnitude improvement over existing ideal-fluid models. 
\end{abstract}

\section{Introduction} 

\begin{figure}[ht]
\centering
\includegraphics[width=0.48\textwidth]{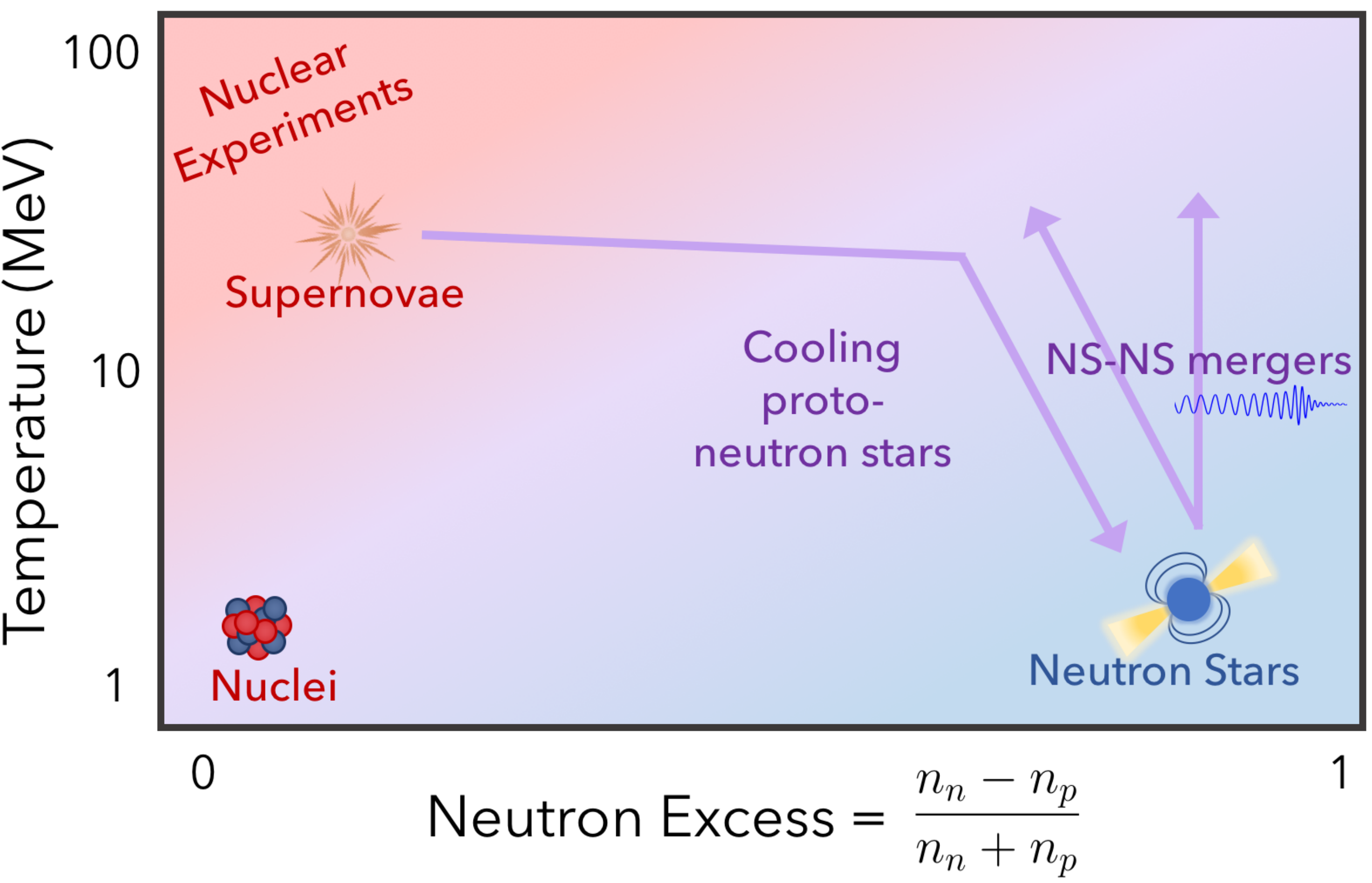}
\caption{\label{fig:phase} Cross-section of a phase diagram, containing temperature as a function of neutron excess, where neutron excess is defined as the difference between neutron and proton densities, $n_n$ and $n_p$, compared to the total baryon density. The approximate regimes probed by various terrestrial and astrophysical phenomena are indicated. The dense-matter EOS is primarily constrained by observations of neutron stars and by laboratory data from nuclei and nuclear experiments. Many dynamical phenomena, such as neutron star mergers, supernovae, and the cooling of proto-neutron stars, lie in the intermediate regions of parameter space where the temperature is non-zero and the matter can be at a variable proton fraction.} 
\end{figure}

Many dynamical phenomena, including core collapse supernovae, the formation and subsequent cooling of proto-neutron stars, and both the electromagnetic and gravitational signals from neutron star mergers, depend sensitively on the neutron star equation of state (EOS) at densities where the EOS is not well understood. In addition, for these dynamical phenomena, there are two further complications. First, temperatures may range from below the Fermi temperature, for which ``cold" EOS suffice, to temperatures of up to 10-100~MeV in neutron star mergers (e.g., \citealt{Oechslin2007}). Second, the composition may range from nearly pure neutron matter to symmetric matter, with some dynamical timescales shorter than the timescale required to establish $\beta$-equilibrium. While astrophysical observations of stationary neutron stars probe the cold EOS in $\beta$-equilibrium and laboratory experiments constrain the hot EOS of symmetric matter, extrapolations between the two regimes remain difficult. (For a schematic representation of these various regimes, see Fig.~\ref{fig:phase}. For recent reviews, see e.g., \citealt{Lattimer2016, Ozel2016}.) Such extrapolations to arbitrary proton fraction and temperature add further uncertainty to the EOS and complicate numerical simulations of these phenomena. 

In the zero-temperature limit, a large number of EOS have been calculated, ranging from purely nucleonic models (e.g., \citealt{Baym1971, Friedman1981, Akmal1998, Douchin2001}) to models incorporating quark degrees of freedom using state-of-the-art results from perturbative QCD \citep[e.g.,][]{Fraga2014}. Laboratory experiments and neutron-star observations do not yet have sufficient power to distinguish between these models. Furthermore, it is likely that these EOS do not span the full range of possible physics. This possibility has motivated the creation of a large number of parametric EOS, as were first introduced by \citet{Read2009} and \citet{Ozel2009}. These parametric models do not require a priori knowledge of the high-density nuclear physics governing the EOS and, hence, can be used to probe unknown physics from neutron star observations.

A much smaller number of EOS that self-consistently incorporate finite-temperature effects have been calculated to date. Among the most well-known of these are the LS model, which is based on finite-temperature compressible liquid drop theory with a Skyrme nuclear force \citep{Lattimer1991}; as well as the EOS of \citet{Shen1998a}, which was calculated using relativistic mean field (RMF) theory with a Thomas-Fermi approximation. More recently, the statistical model developed in \citet{Hempel2010} has been applied to an additional $\sim$10 combinations of RMF models and nuclear mass tables.

Just as parametrizations of the cold EOS have proven useful in representing a broader range of physics, so too would a parametric finite-temperature EOS be useful for incorporating EOS effects into supernova and merger calculations. To this end, many authors have employed so-called ``hybrid EOS," in which a thermal component for an ideal fluid is added to an arbitrary cold EOS to account for heating \citep{Janka1993}. The ideal-fluid thermal component is parametrized in terms of a simple adiabatic index as $P_{\rm th} = \epsilon_{\rm th} (\Gamma_{\rm th}-1)$, where $P_{\rm th}$ and $\epsilon_{\rm th}$ are the thermal pressure and energy density and $\Gamma_{\rm th}$ is the adiabatic index, which is assumed to be constant. Such an approach is computationally simple, but neglects the effect of degeneracy on the thermal pressure. At high densities and finite temperatures, part of the available energy acts to lift degeneracy, rather than contributing additional thermal support. This causes a net reduction in the thermal pressure at high densities, compared to the prediction for an ideal fluid.

The density-dependence of these thermal effects depends directly on the density-dependence of the nucleon effective mass, as has been shown for many EOS \citep{Constantinou2014,Constantinou2015}. \citet{Constantinou2015} performed a Sommerfeld expansion to approximate the thermal properties at next-to-leading order and showed that the expansion terms require both the effective mass and its derivatives. Given a complete expression for the density-dependence of the effective mass, they showed that this formalism can be used to accurately approximate the thermal properties of a wide variety of EOS. \citet{Constantinou2017a} later expanded this work and showed that the formalism can be used to recreate even models beyond mean field theory, such as the two-loop exchange model of \citet{Zhang2016}. 

The strong dependence of thermal properties on the effective mass can also be seen in the behavior $\Gamma_{\rm th}$. For example, \citet{Constantinou2015}  compared two EOS with similar zero-temperature properties but with different single-particle potentials, and hence different density-dependences in their nucleon effective masses. They found substantially different thermal properties for the two EOS and that a constant $\Gamma_{\rm th}$ model failed to describe either EOS. \citet{Zhang2016} also found a strong density-dependence in $\Gamma_{\rm th}$ for their two-loop exchange model. These results indicate that $\Gamma_{\rm th}$ has a significant density-dependence for a diverse range of analytic models, which is not captured in the constant $\Gamma_{\rm th}$ approximation of the hybrid EOS.

Neglecting the effect of degeneracy on the thermal pressure has important consequences for dynamical simulations as well. For example, \citet{Bauswein2010} compared the properties of a neutron star-neutron star merger that would be predicted by a hybrid EOS and by more realistic EOS. Specifically, they compared the \citet{Shen1998a} and \citet{Lattimer1991} EOS to hybrid EOS that were constructed from the zero-temperature versions of these same EOS with either $\Gamma_{\rm th}=1.5$ or 2. They found that using the hybrid EOS predicts post-merger frequencies from a hypermassive neutron star that are 50-250~Hz smaller than what is found with a realistic finite-temperature EOS. Moreover, the lifetime of the hypermassive remnant can deviate by a factor of two from the more realistic value and the post-collapse accretion disk mass around the resulting black hole can differ by up to 30\% when the simplified thermal effects are used \citep{Bauswein2010}. These results all suggest that it is indeed important to account for the effect of degeneracy on the thermal pressure when simulating neutron star mergers.

The Sommerfeld expansion results of \citet{Constantinou2015} can be used to explicitly correct a hybrid EOS to include degenerate effects, as long as the particle interactions and potentials of the cold EOS are known. However, requiring knowledge of the potentials of the cold EOS renders these corrections inapplicable to piecewise-polytropic EOS or other parametric forms of the EOS that are agnostic in their descriptions of the microphysics.

The goal of this paper is to develop a physically-motivated framework for incorporating the thermal pressure that maintains the wide applicability of the hybrid EOS approach. With such a model, it will be possible to robustly add thermal effects to any cold EOS in $\beta$-equilibrium, without having to make the simplifying assumptions of an ideal fluid at all densities. The framework we present in this paper is specific to neutron-proton-electron (\npe) matter, but could be generalized to include more exotic particles. We also include a symmetry-energy dependent correction that extrapolates the proton fraction away from $\beta$-equilibrium. The complete model thus allows us to build an EOS at finite-temperature and arbitrary proton fraction from any cold \npe~EOS in neutrinoless $\beta$-equilibrium, including piecewise-polytropic EOS. Moreover, the model is analytic and in closed-form and thus can be calculated efficiently in dynamical simulations.

We start in $\S$\ref{sec:overview} with a brief review of existing finite-temperature EOS and a discussion of the regimes in which thermal effects become important. In $\S$\ref{sec:outline}, we outline our model. We provide the symmetry-energy dependent extrapolation to arbitrary proton fraction in $\S$\ref{sec:Esym}. In $\S$\ref{sec:thermal}, we introduce our $M^*$-approximation of the thermal effects. We summarize the model in $\S$\ref{sec:boxes}, in which all of the relevant equations can be found in Boxes I and II. Finally, we quantify the performance of our model in $\S$\ref{sec:complete}. We find that with a relatively small set of parameters, our complete model is able to recreate existing finite-temperature EOS with introduced errors of $\lesssim$20\%, for densities above the nuclear saturation density. 

\section{Overview of finite-temperature EOS}
\label{sec:overview}
Before introducing our new approximation for the pressure at arbitrary proton fraction and temperature, we will first briefly review the finite-temperature EOS that have been previously developed. 

Two of the most widely-used finite-temperature EOS are the models of \citet[][hereafter LS]{Lattimer1991}, which is based on a finite-temperature liquid drop model with a Skyrme nuclear force, and \citet[][hereafter STOS]{Shen1998a} which is an RMF model that is extended with the Thomas-Fermi approximation. An additional eight EOS have been calculated with the framework of \citet[][hereafter, HS]{Hempel2010}, which is a statistical model that consists of an ensemble of nuclei and interacting nucleons in nuclear statistical equilibrium and, hence, goes beyond the single nucleus approximation that both LS and STOS assume. Each HS EOS represents the nucleons with an RMF model and additionally includes excluded volume effects. Of the RMF models that have been used with the HS method, six are nucleonic: TMA \citep{Toki1995}, TM1 \citep{Sugahara1994}, NL3 \citep{Lalazissis1997}, FSUGold \citep{Todd-Rutel2005}, IUFSU \citep{Fattoyev2010}, DD2 \citep{Typel2010}; while the models BHB$\Lambda \phi$ and BHB$\Lambda$ include hyperons with and without the repulsive hyperon-hyperon interaction mediated by the $\phi$ meson, respectively \citep{Banik2014}. Additionally, \citet{Steiner2013} created a set of two finite-temperature EOS, SFHo/x, that also used the statistical method of HS, but with new RMF parameterizations and constraints from neutron star observations. There are also the EOS of G. Shen, which are based on a virial expansion and nuclear statistical equilibrium calculations at low densities and RMF calculations at high densities, using the models FSUGold \citep{Shen2011} and NL3 \citep{Shen2011a}. Tables of these various EOS can be found on the website of M. Hempel,\footnote{https://astro.physik.unibas.ch/people/matthias-hempel/equations-of-state.html} \texttt{stellarcollapse.org}, and/or the \texttt{CompOSE} database.\footnote{https://compose.obspm.fr/home/}

More recently, several new finite-temperature EOS have been added to the \texttt{CompOSE} database. These include the SLY4-RG model, which is calculated in nuclear statistical equilibrium using a Skyrme energy functional \citep{Gulminelli2015,Raduta2018}, as well as chiral mean field theory models, which include hyperons as additional degrees of freedom \citep[e.g.,][]{Dexheimer2017}, generalized relativistic density functional models \citep[e.g.,][]{Typel2018}, and models calculated using a variational method applied to two- and three-body nuclear potentials \citep[e.g.,][]{Togashi2017}.

For the sake of simplicity in the following analysis, we will focus on a subset of these EOS and will include only models that are nucleonic. In particular, our sample will include STOS as well as the eight nucleonic EOS calculated with the HS method, to represent the models based on RMF theory. We will also include LS (with a compression modulus $K=220$~MeV) and SLY4-RG, to represent non-relativistic models with Skyrme nuclear forces.

In spite of the increasing number of finite-temperature EOS that have been calculated, they nevertheless span a relatively limited range of physics, especially when compared to the diversity of cold EOS models. In order to span a broader range of possible physics, many authors have used the so-called ``hybrid EOS," which assume that the thermal pressure is given simply by an ideal-fluid term that can be added to any cold EOS. The hybrid EOS were first introduced by \citet{Janka1993} and have been used in many subsequent works \citep[for recent reviews, see][]{Shibata2011, Faber2012,Baiotti2017, Paschalidis2017}. In these hybrid EOS, the thermal pressure is written as
\begin{equation}
\label{eq:Pthhyb}
P_{\rm th, hybrid}(n,T) = n E_{\rm th, hybrid}(n,T)  ( \Gamma_{\rm th} - 1),
\end{equation}
where $E_{\rm th,hybrid}(n,T)$ is the thermal contribution to the energy per baryon, $n$ is the baryon number density, and $\Gamma_{\rm th}$ is the thermal adiabatic index and is constrained to be $1 \le \Gamma_{\rm th} \le 2$. In the hybrid approximation, $\Gamma_{\rm th}$ is assumed to be constant.

Following \citet{Etienne2008}, the hybrid temperature-dependence of $E_{\rm th,hybrid}$ is included as an ideal fluid plus a contribution from relativistic particles, i.e.,
\begin{equation}
\label{eq:Ethhyb}
E_{\rm th,hybrid}(n,T) = \frac{3}{2} k_B T + \frac{4 \sigma}{c}\frac{f_s}{ n} T^4,
\end{equation}
where $k_B$ is the Boltzmann constant, $T$ is the temperature, and $\sigma \equiv \pi^2 k_B^4 / [60 \hbar^3 c^2]$ is the Stefan-Boltzmann constant, with $\hbar$ the Planck constant and $c$ the speed of light. The parameter $f_S$ represents the number of ultra-relativistic species that contribute to the thermal pressure. For $k_BT\ll 2 m_e c^2$, where $m_e$ is the mass of an electron, photons will dominate and $f_S$=1. For $k_BT\gg 2 m_e c^2$, electrons and positrons become relativistic as well and yield $f_S = 1 + 2\times (7/8) = 11/4$. Finally, for $k_BT \gtrsim 10$~MeV, thermal neutrinos and anti-neutrinos appear, rendering $ f_s = 11/4 +  3 \times (7/8) = 43/8$. If right-handed neutrinos were to exist, this would become $f_s = 11/4+ 3\times2\times(7/8) = 8$. 

We note that all 12 EOS discussed above neglect neutrinos in their calculations. The STOS EOS additionally neglects leptons and photons, which we add in wherever we use STOS in this paper. For the STOS thermal lepton and photon contribution, we use eq.~(\ref{eq:Ethhyb}) with the appropriate lepton density. For the cold lepton energy, we add the contribution for a degenerate gas of relativistic electrons. Because all the EOS neglect neutrinos, we will also neglect neutrinos in our comparisons and thus we will calculate $f_S$ only as 
\begin{equation}
\label{eq:fs}
 f_S = 
\begin{cases}
1,   & k_B T < 1~\text{MeV} , \\
 11/4,  & k_B T \ge~1 \text{MeV}.
\end{cases}
\end{equation}

We, therefore, account for the degrees of freedom introduced by the possible presence of ultra-relativistic positrons. However, throughout this paper, we will assume that the population of positrons is small and that their contribution to the pressure or energy at higher densities is negligible. If there were a scenario in which the population of positrons were significant compared to the electrons, one would have to explicitly account for the positrons in particle-counting as well as in imposing charge neutrality.

\begin{figure}[ht]
\centering
\includegraphics[width=0.48\textwidth]{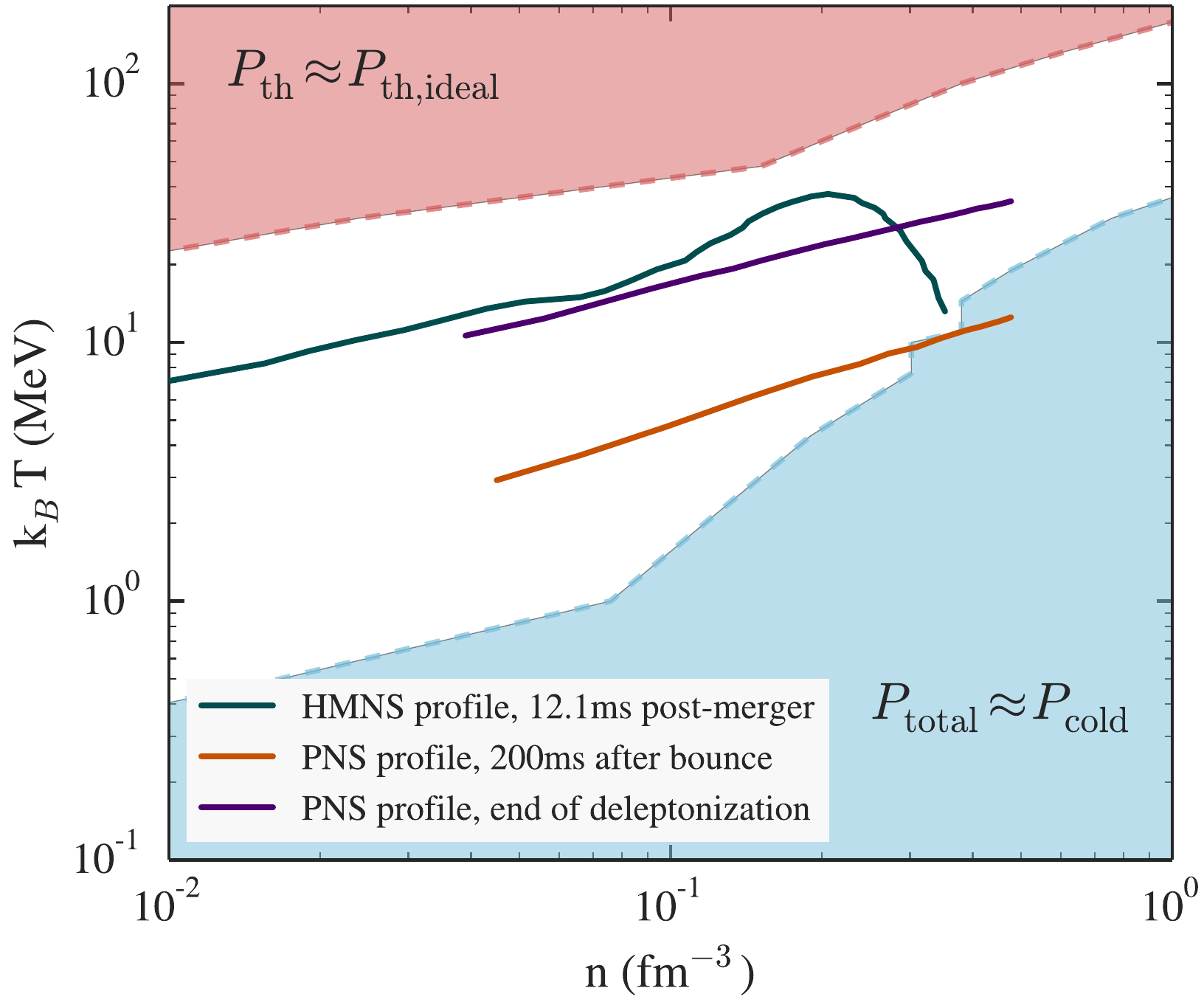}
\caption{\label{fig:Tn} Phase diagram for regimes of interest in neutron star simulations. The blue shaded region represents the regime where the total pressure is dominated by the cold pressure, to within $1\%$, for the STOS EOS with proton fraction  $Y_p=0.1$. The red shaded region represents the $T-n$ range where the thermal pressure is dominated by the ideal-fluid pressure ($P_{\rm th}=nk_B T$), to within $1\%$, for the same EOS and fixed $Y_p$. The white range in between these two extremes represents the phase space in which degenerate thermal effects are important. For comparison, the green line shows the profile of a hypermassive neutron star (HMNS) remnant 12.1~ms after a neutron star merger from the simulations of \citet{Sekiguchi2011} using the STOS EOS. The orange and purple lines show the profiles of a proto-neutron star (PNS) 200~ms after the bounce in a core-collapse supernova simulation and at the end of de-leptonization in the same simulation, both with a bulk version of the LS EOS \citep{Camelio2017}.  } 
\end{figure}

In order to highlight the regimes where a realistic finite-temperature EOS and the hybrid approximation differ, we show a phase diagram in Fig.~\ref{fig:Tn}.  In this plot, we show various regions calculated for the EOS STOS, all at a fixed proton fraction of $Y_p=0.1$. The total pressure, $P_{\rm total}$, is thus calculated at $Y_p=0.1$ and a given temperature. The cold contribution, $P_{\rm cold}$, is calculated at the same $Y_p$ and at zero-temperature.\footnote{We note that, throughout this paper, we use the coldest HS calculation, performed at $k_BT=0.1$~MeV, as an approximation of the zero-temperature EOS. Even though the STOS EOS is calculated at $T=0$~MeV, we use the $k_BT=0.1$~MeV table as our cold component for this EOS as well, in order to maintain consistency with the HS set of EOS.} Finally, the thermal contribution, $P_{\rm th}$, is defined as $P_{\rm total} -P_{\rm cold}$ for the same proton fraction.

In this figure, the blue shaded region shows the regime where the total pressure is dominated by the cold pressure; there, the thermal pressure of STOS contributes $<1\%$ of the total pressure. The red shaded region represents the regime where the thermal pressure can be approximated by the ideal fluid pressure ($P_{\rm th, ideal}=nK_BT$), to within 1\%. The white region between these two extremes represents the range of parameter space in which the thermal pressure is important but the ideal-fluid approximation does not yet apply. In this white region, the effects of degeneracy on the thermal pressure cannot be neglected. 

For comparison, we also show in Fig.~\ref{fig:Tn} the projected temperature-density profiles from three different simulations of relevant astrophysical phenomena. The green line shows the profile of a hypermassive neutron star remnant 12.1~ms after the merger of two 1.35~$\Ms$ neutron stars, as simulated using the EOS STOS \citep{Sekiguchi2011}. The orange and purple lines both come from numerical simulations of the evolution of a proto-neutron star using a bulk-version of the LS EOS. The orange line gives the profile of the proto-neutron star at 200~ms after the core bounce, while the purple line shows the profile of the proto-neutron star at the end of the de-leptonization phase \citep{Camelio2017}. We note that these profiles are not necessarily calculated at $Y_p=0.1$, but we include them nevertheless to show the approximate relevant temperatures and densities for such phenomena.

In order to further explore the dependence on the proton fraction, we also calculated the regime where degeneracy dominates for increasing values of $Y_p$. We find that as the proton fraction increases towards $Y_p=0.5$, the white degeneracy region in Fig.~\ref{fig:Tn} shrinks, but still largely encompasses the shown profiles. We thus find that all of these simulations primarily probe the phase space where degenerate thermal effects are important. This suggests that using the hybrid approximation, instead of the full thermal pressure, may bias the outcomes from such simulations.

\section{Generic model of a finite temperature EOS}
\label{sec:outline}
In order to construct a finite-temperature EOS at arbitrary proton fraction, our model must be able to extrapolate from $\beta$-equilibrium to an arbitrary $Y_p$, as well as from cold matter to an arbitrary temperature. This will naturally introduce dials into our model that can be adjusted to represent a wide range of physics, based on the symmetry energy, its slope, and the strength of particle interactions we wish to include. Moreover, we will show that with a small set of parameters, the EOS that are currently in use in the literature can be replicated to high accuracy.

We start with our model in general terms, for which we will derive analytic expressions in the following sections. Our final model will be for the complete energy per baryon, $E(n,Y_p,T)$, separated into analytic, physically-motivated terms. A summary of the final equations can be found in Boxes I and II in $\S$\ref{sec:boxes}.

We can expand the energy per particle of nuclear matter, $E_{\rm nucl}$, about the neutron excess parameter, $(1-2 Y_p)$, to second order as
\begin{equation}
\label{eq:parabolic}
E_{\rm nucl}(n, Y_p, T) = E_{\rm nucl}(n, Y_p=\sfrac{1}{2}, T) + E_{\rm sym}(n, T)(1-2 Y_p)^2,
\end{equation}
where $E_{\rm nucl}(n, Y_p = \sfrac{1}{2}, T)$ represents the energy of symmetric nuclear matter and
\begin{equation}
E_{\rm sym}(n,T) \equiv \frac{1}{2} \frac{\partial^2 E_{\rm nucl}(n, Y_p, T)}{\partial (1-2 Y_p)^2} \biggr|_{Y_p=1/2}
\end{equation}
is the symmetry energy.  The proton fraction is related to the overall baryon density, $n$, according to
\begin{equation}
\label{eq:Yp}
Y_p = \frac{n_p}{n} = \frac{N_p}{N_n + N_p},
\end{equation}
where $n_p$ is the proton density, $N_p$ is the total number of protons, and $N_n$ is the total number of neutrons. Throughout this paper, we enforce charge-neutrality, which requires that the proton and electron densities balance. Thus, the electron density, $n_e$, can be written as
\begin{equation}
\label{eq:ne}
n_e = Y_p n.
\end{equation}
Finally, by requiring that the baryonic components combine to give the total density $n$, we can write the neutron density as
\begin{equation}
n_n = (1-Y_p) n.
\end{equation}

We can further expand eq.~(\ref{eq:parabolic}) by separating the energy of cold, symmetric matter from its thermal contribution, i.e., 
\begin{align}
\begin{split}
\label{eq:intermidateE}
E_{\rm nucl}(n, Y_p, T) = & E_{\rm nucl}(n, Y_p=\sfrac{1}{2}, T=0)  \\
	 + & E_{\rm nucl,th}(n, Y_p=\sfrac{1}{2},T) \\
	 + & E_{\rm sym}(n, T)(1-2 Y_p)^2.
\end{split}
\end{align}
Here and throughout the paper, we use the subscript ``\textit{th}" to indicate the thermal contribution to a variable, after the cold component has been subtracted.

In order to write the energy with respect to a cold EOS in $\beta$-equilibrium, as is often most relevant to start from in the study of neutron stars, we eliminate the cold, symmetric term in eq.~(\ref{eq:intermidateE}) to yield
\begin{align}
\begin{split}
E_{\rm nucl}(n, Y_p, T) = & E_{\rm nucl}(n, Y_{p,\beta}, T=0)  \\
	 + & E_{\rm nucl,th}(n, Y_p=\sfrac{1}{2},T) \\
	 + & E_{\rm sym}(n, T)(1-2 Y_p)^2 \\
	 - & E_{\rm sym}(n, T=0)(1-2 Y_{p,\beta})^2,
\end{split}
\end{align}
where $Y_{p,\beta}$ represents the proton fraction of a zero-temperature system in $\beta$-equilibrium. We note that the proton fraction depends on the density, i.e., $Y_{p,\beta} =Y_{p,\beta}(n)$, but for simplicity we suppress this in our notation.

Finally, we must add the contribution of leptons and photons to this expression. The zero-temperature energy from relativistic degenerate electrons is given by
\begin{equation}
E_{\rm lepton}(n, Y_p, T=0) =  3 K Y_p (Y_p n)^{1/3},
\end{equation}
where the extra factor of $Y_p$ comes from our definition of $E$ as the energy per baryon, combined with eqs.~(\ref{eq:Yp}) and (\ref{eq:ne}). Here, $K\equiv (3 \pi^2)^{1/3} ( \hbar c/4)$. Additionally, there will also be a thermal contribution, $E_{\rm lepton,th}(n,Y_p,T)$, which we derive in $\S$\ref{sec:thermal}.

Thus, our skeletal model for the total energy is given by the following set of equations:

\begin{subequations}
\begin{align}
\begin{split}
\label{eq:fullE}
&E(n, Y_p, T)   = E(n, Y_p, T=0) + E_{\rm th}(n, Y_p, T)
\end{split} \\
\begin{split}
\label{eq:Ecold}
\\& E(n, Y_p, T=0)  = E(n, Y_{p,\beta}, T=0)  \\
	& \quad  + E_{\rm sym}(n, T=0)\left[(1-2 Y_p)^2 - (1-2 Y_{p,\beta})^2\right] \\
	& \quad + 3 K  \left( Y_p^{4/3} - Y_{p,\beta}^{4/3} \right) n^{1/3}  
\end{split} \\
\begin{split}
\label{eq:Eth}
& E_{\rm th}(n, Y_p, T)  = E_{\rm nucl,th}(n, Y_p=\sfrac{1}{2},T) \\
	& \qquad + E_{\rm lepton,~ th}(n, Y_p, T) \\
	& \qquad + E_{\rm sym,th}(n, T)(1-2 Y_p)^2 .
\end{split}
\end{align}
\end{subequations}

From these relations, we can derive the pressure via the standard thermodynamic relation,
\begin{equation}
\label{eq:getP}
P \equiv -\frac{\partial U}{\partial V}\biggr \rvert_{N_q,S} = n^2 \left[ \frac{\partial E(n,T=0)}{\partial n}\right] \biggr \rvert_{Y_p, S}
\end{equation}
where $U$ is the total energy, $V$ is the volume, $N_q$ is the number of each species $q$, and $S$ is the total entropy. From eq.~(\ref{eq:Yp}), it is clear that evaluating these derivatives at constant $N_q$ is equivalent to evaluating them at constant $Y_p$. In this paper, we will mainly plot results in terms of pressure. We summarize the complete expressions for pressure in Box II of \S\ref{sec:boxes}.

While this set of expressions may seem to have a large number of terms, this separation allows these terms to be represented analytically. Moreover, as we will show, the parameters of each term are linked directly to physics on which there  are experimental constraints and of which further constraints are the motivation of many observations of astrophysical neutron stars:  namely, the value of the symmetry energy at the saturation density, the slope of the symmetry energy, and the strength of interactions between particles. 

\section{Derivation of the cold symmetry energy in the Fermi Gas limit}
\label{sec:Esym}
We turn first to the symmetry energy correction term, $E_{\rm sym}(n, T)$ of eq.~(\ref{eq:parabolic}). The symmetry energy is defined as the per-nucleon difference in energy between symmetric matter and pure neutron matter. In other words, the symmetry energy represents the excess energy of matter with unequal numbers of protons and neutrons. In nuclear models, the symmetry energy is typically calculated as an expansion around the nuclear saturation density, for matter with $Y_p=1/2$. In eq.~(\ref{eq:parabolic}), we perform the expansion with respect to the proton fraction and, in the following section, will introduce a density-dependence to extrapolate beyond the saturation density, where the coefficients of our approximation are experimentally constrained.  In this section, we will provide the approximation for $E_{\rm sym}(n,T)$ at zero-temperature. For the thermal contribution to the symmetry energy, which turns out to be negligible, see $\S$\ref{sec:thermal}.

It is particularly useful to parameterize the symmetry energy in terms of its separate kinetic and potential components at zero-temperature \citep[e.g.,][]{Tsang2009, Steiner2010}, modified by a parameter $\eta$ to account for short-range correlations due to the tensor force acting between a spin-triplet or isospin-singlet proton-neutron pair. These correlations can significantly reduce the kinetic symmetry energy to even a negative value at the saturation density, compared to the kinetic energy of an uncorrelated Fermi gas model \citep{Xu2011, Vidana2011, Lovato2011, Carbone2012, Rios2014,Hen2015}. In this framework, we parameterize the symmetry energy of eq.~(\ref{eq:Ecold}) as
\begin{equation}
\label{eq:Esym}
E_{\rm sym}(n, T=0)= \eta E_{\rm sym}^{\rm kin}(n)  
	+ \left[ S_0 - \eta E_{\rm sym}^{\rm kin}(n_{\rm sat}) \right] \left(\frac{n}{n_{\rm sat}}\right)^{\gamma},
\end{equation}
as in \citet{Li2015}. Here, $E_{\rm sym}^{\rm kin}(n)$ is the ``kinetic" symmetry energy, arising from the change in the Fermi energy of a gas at density $n$ as the relative proton/neutron fraction changes, $n_{\rm sat}=0.16$~fm$^{-3}$ is the nuclear saturation density,\footnote{We note that $n_{\rm sat}$ does vary slightly among the EOS in our sample, but we fix the value to $n_{\rm sat}=0.16$~fm$^{-3}$ in order to more easily compare the various EOS. We find that this does not significantly affect the results.}  and the second term represents the ``potential" symmetry energy which accounts for the interactions between particles. Because the exact form of the potential symmetry energy is not well known, it is anchored at the saturation density by the magnitude of the overall symmetry energy, $S_0 \equiv E_{\rm sym}(n_{\rm sat})$, and is given an arbitrary density-dependence through the constant $\gamma$. 

In contrast, the kinetic energy term can be calculated directly from the nuclear momentum distribution. The kinetic energy of a free Fermi gas is given simply by
\begin{equation}
\label{eq:fermiKE}
\frac{\varepsilon_{k,q}}{n} = \frac{3}{5} E_f(n_q)  
\end{equation}
where $\varepsilon_{k,q}$ is the kinetic energy per particle, $q$ represents the particle (either a neutron or proton), and $E_f(n)$ is the Fermi energy,
\begin{equation}
 E_f(n_q) = \frac{\hbar^2}{2 m} \left( 3\pi^2 n_q \right)^{2/3},
\end{equation}
in which $m$ is the mass of the relevant particle. For our approximation, we will neglect the small difference between the proton and neutron mass and simply take $m\approx m_n$, where $m_n$ is the neutron mass.

By taking the difference between symmetric matter and pure neutron matter, the kinetic symmetry energy as a function of the total density is then
\begin{align}
\begin{split}
\label{eq:Ekinsym}
E_{\rm sym}^{\rm kin}(n)  &= \frac{3}{5} \left[ 2E_f\left(n_p = n_n =\frac{1}{2} n\right) - E_f(n_n = n)  \right]  \\
 &= \frac{3}{5} \left(2^{1/3} -1 \right)  E_f(n).
\end{split}
\end{align}

We can also eliminate the parameter $\eta$ in eq.~(\ref{eq:Esym}) by introducing the constant $L$, which is related to the overall slope at the saturation density via,
\begin{equation}
\label{eq:L}
 L \equiv  3  n_{\rm sat} \left[ \frac{\partial E_{\rm sym}(n, T=0)}{\partial n} \right]\biggr\rvert_{n_{\rm sat}}.
\end{equation}
Combining eqs.~(\ref{eq:Esym}) and (\ref{eq:L}), we can solve for $\eta$ in terms of the quantities $S_0$ and $L$, which are constrained by nuclear physics experiments for matter near $Y_p=1/2$ \citep{Lattimer2013}. We find
\begin{equation}
\label{eq:eta}
\eta = \frac{5}{9} \left[ \frac{ L-3 S_0 \gamma}{\left(2^{1/3}-1\right)\left(2/3 - \gamma \right) E_f(n_{\rm sat})} \right]  ,
\end{equation}
thereby leaving one free parameter, $\gamma$, which is constrained by nuclear experiments to lie in the range $\sim$~0.2 to 1.2 (see, e.g., Fig.\ 2 of \citealt{Li2015}; \citealt{Tsang2009}).

\begin{deluxetable}{llll}
\tabletypesize{\footnotesize}
\tablewidth{0.48\textwidth} 
\tablecaption{\label{table:gamma} Symmetry energy parameters characterizing each EOS at $k_BT=0.1$~MeV.  }
\tablehead{\\
\colhead{ EOS }  &
\colhead{$S_0$ (MeV) }  &
\colhead{$L$ (MeV) }  &
\colhead{$\gamma$ }  
}
\startdata
          TM1 &  36.95 & 110.99  & 0.75 \\
          TMA &  30.66 & 90.14  & 0.66 \\
          NL3 &  37.39 & 118.49  & 0.62 \\
          FSG &  32.56 & 60.43  & 1.11 \\
          IUF &  31.29 & 47.20  & 0.52 \\
          DD2 &  31.67 & 55.03  & 0.91 \\
        STOS &  36.95 & 110.99  & 0.77 \\
         SFHo &  31.57 & 47.10  & 0.41 \\
         SFHx &  28.67 & 23.18  & -0.04\footnote{The inferred value for $\gamma$ for SFHx is highly sensitive to the density range that is included in the fit; see the discussion in the text for details.} \\
	 LS &  29.3 &  74.0	&  1.05 \\
 	SLY4-RG &	 32.04	&	46.00 &	0.35 

\enddata
\tablecomments{$S_0$ and $L$ are fixed to the values predicted for each EOS, while $\gamma$ is a fit parameter. All fits are performed for densities above $n\ge0.01$~fm$^{-3}$ and $n_{\rm sat}$=0.16~fm$^{-3}$}.
\end{deluxetable}

We thus have a complete expression for the symmetry energy that depends only on the three parameters $\gamma$, $S_0$, and $L$ which, in principle, can be constrained by nuclear experiments. We can now use this functional form to fit for $\gamma$, by combining it with the following relationship between the symmetry energy and $Y_{p,\beta}$ for  charge-neutral \npe~matter in neutrinoless $\beta$-equilibrium,

\begin{equation}
\label{eq:YpBeta}
\frac{Y_{p,\beta}}{(1-2 Y_{p,\beta})^3} = \frac{64}{3 \pi^2 n} \left[ \frac{E_{\rm sym}(n, T=0)}{ \hbar c}\right]^3
\end{equation}
\citep[for a derivation of this relation, see, e.g.,][or Appendix A]{Blaschke2016}.  When solved for $Y_{p,\beta}$, this becomes
\begin{equation}
\label{eq:YpBInv}
Y_{p,\beta} = \frac{1}{2} + \frac{(2 \pi^2)^{1/3}}{32} \frac{n}{\xi} \left\{ (2\pi^2)^{1/3} - \frac{\xi^2}{n} \left[\frac{\hbar c}{E_{\rm sym}(n,T=0)}\right]^3 \right\},
\end{equation}
where, for simplicity, we have introduced the auxilary quantity $\xi$, defined as
\begin{multline}
\label{eq:xi}
\xi \equiv \left[ \frac{E_{\rm sym}(n,T=0)}{\hbar c} \right]^2  \times \\
\left\{ 24 n \left[ 1+ \sqrt{ 1 +  \frac{\pi^2 n}{288}\left(\frac{\hbar c}{E_{\rm sym}(n,T=0) }\right)^3}\right]  \right\}^{1/3}.
\end{multline}

For each of the EOS in our sample, we stitch together a complete cold EOS at $\beta$-equilibrium from the publically-available tables at fixed $Y_p$, by requiring that $\mu_e + \mu_p - \mu_n = 0$, where $\mu_i$ is the chemical potential of each species. We then use the corresponding density-dependent proton fraction, $Y_{p,\beta}$, to fit for $\gamma$ using eqs.~(\ref{eq:Esym})-(\ref{eq:YpBeta}) and keeping $S_0$ and $L$ fixed for each EOS. We perform the fits using a standard least-squares method and limit the density range to $n \ge 10^{-2}$~fm$^{-3}$. In principle, eqs.~(\ref{eq:Esym})-(\ref{eq:YpBeta}) apply only to \npe~matter, which will be uniform only above $0.5 n_{\rm sat}$. However, in practice, we find a very small difference in the fits for $\gamma$ whether we include densities above $0.5n_{\rm sat}=0.08$~fm$^{-3}$ or whether we start the fits at a slightly lower but still astrophysically relevant cutoff of $n=10^{-2}$~fm$^{-3}$. We show the resulting fit values in Table~\ref{table:gamma}. 

We note that the range of EOS provided in Table~\ref{table:gamma} is intentionally broad. While the symmetry energy parameters of some of these EOS disagree with the combined set of experimental constraints (see \citealt{Lattimer2013} for a recent review), or are in disagreement with certain theoretical considerations such as chiral effective field theory results for pure neutron matter \citep[see, e.g.,][]{Kruger2013}, they are all consistent with at least some experimental constraints on $S_0$ and $L$.

We find that $\gamma$ spans roughly the range of experimentally-allowed values, between 0.15 and 1.0, as expected, with the exception of SFHx. SFHx has an extremely low value of $L$, which makes the result of the fit highly sensitive to the density range that is included. For consistency, we still constrain the densities to $n \ge 10^{-2}$~fm$^{-3}$ for the fit to this EOS; however, the inferred value for $\gamma$ ranges from the reported value of $-0.04$ up to 0.18, depending on where the density cutoff is placed. Thus, the particular value for $\gamma$ for SFHx should be taken with some caution.

We have here used eq.~(\ref{eq:YpBeta}) to fit for $\gamma$ from the $\beta$-equilibrium proton fractions of realistic EOS. We wish to also emphasize that eq.~(\ref{eq:YpBeta}) can, of course, be used to calculate $Y_{p,\beta}$, given a choice of $S_0, L$, and $\gamma$. Once these three parameters are specified, eqs.~(\ref{eq:YpBInv})-(\ref{eq:xi}) can be used to calculate $Y_{p,\beta}$ for any EOS. As a result, all that is required of the cold EOS is knowledge of the run of pressure with density. This feature makes it possible to apply our model to piecewise polytropes or other families of parametric EOS that may not directly calculate $Y_{p,\beta}$.

\begin{figure}[ht]
\centering
\includegraphics[width=0.47\textwidth]{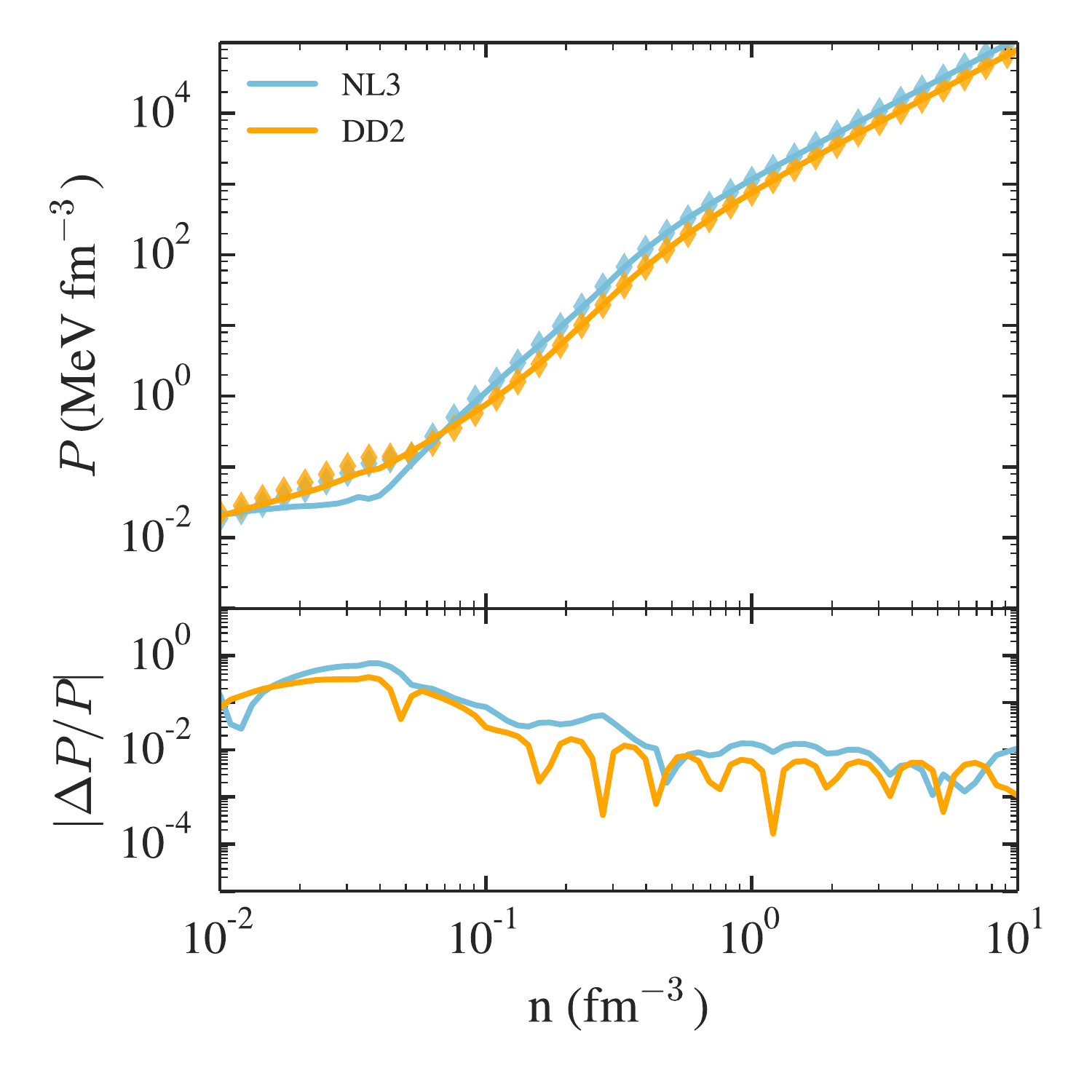}
\caption{\label{fig:coldYpcorr} Top: Pressure as a function of density for EOS NL3 and DD2, at $k_BT=0.1$~MeV and $Y_p=0.1$, as blue and orange diamonds, respectively. The solid lines show our model of the pressure, calculated using eqs.~(\ref{eq:Ecold}) and (\ref{eq:Esym}-\ref{eq:YpBeta}). Our model starts with the respective EOS in $\beta$-equilibrium and adds the appropriate symmetry energy and lepton corrections to extrapolate to $Y_p=0.1$. For $S_0, L$, and $\gamma$, we use the values listed in Table~\ref{table:gamma}. Bottom: Residuals between the true EOS at $Y_p=0.1$ and our model. We find that our model extrapolates from $\beta$-equilibrium to $Y_p=0.1$ reasonably well, especially at high densities where the model introduces an error of $\lesssim 1\%$ compared to using the full EOS.} 
\end{figure}

We show an example of the performance of this model for $E_{\rm sym}(n, T=0)$ in Fig.~\ref{fig:coldYpcorr} for the EOS NL3  \citep{Lalazissis1997, Lalazissis1999} and DD2 \citep{Typel2010}. We show these two EOS as representative samples, with NL3 representing the family of EOS with larger $L$ values and DD2 representing the EOS with smaller symmetry energy slopes (see Table~\ref{table:gamma}). The top panel of Fig.~\ref{fig:coldYpcorr} shows the zero-temperature pressure predicted by NL3 and DD2 at $Y_p=0.1$ as blue and orange diamonds, respectively. The colored lines show our model:  starting with the corresponding EOS in $\beta$-equilibrium, adding the symmetry energy correction of eqs.~(\ref{eq:Esym})-(\ref{eq:eta}), and correcting for the leptons, all according to eq.~(\ref{eq:Ecold}). For these models, we take the values of $S_0$, $L$, and $\gamma$ for each EOS from Table~\ref{table:gamma}. We note that we are plotting pressures, but could have similarly shown the energy. We use eq.~(\ref{eq:getP}) to convert the equations of this section to pressures; for the complete set of pressure expressions, see $\S$\ref{sec:boxes} and Box II.

The bottom panel of Fig.~\ref{fig:coldYpcorr} shows the residuals between our model and the pressure predicted by each EOS at $Y_p=0.1$. We find that our model performs very well at densities above $0.5~n_{\rm sat}$, with errors $\lesssim10\%$. At the highest densities, using our model compared to the full EOS introduces errors of only $\sim$1\%. The residuals for the other EOS in our sample are comparably small.

For $Y_p$=0.3, we find the residuals between our model and NL3 and DD2 are comparable to those shown in Fig.~\ref{fig:coldYpcorr}. We, therefore, conclude that this model reasonably captures the $Y_p$-dependence of the cold EOS, for a large range of $L$ values.

We thus have an expression for the symmetry energy at zero-temperature that depends only on $n, Y_p,$ $S_0$, $L$, and the narrowly-constrained parameter $\gamma$. There are two possible routes for creating a finite-temperature EOS with this framework. One possibility is to start from a cold, physically-motivated EOS, which will provide predicted values for $S_0$, $L$, and $Y_{p,\beta}$. In this case, eq.~(\ref{eq:YpBeta}) can be used to fit for $\gamma$. We have provided such fits for the EOS in our sample in Table~\ref{table:gamma}. Alternatively, a cold, parametric EOS can be chosen, for which the underlying physics are not specified. In this case, a user can freely specify $S_0$, $L$, and $\gamma$, which will uniquely specify $Y_{p,\beta}$. For the EOS in our sample, we find that this approach is able to accurately extrapolate from $\beta$-equilibrium to arbitrary proton fraction, introducing errors of $\lesssim10\%$ for densities of interest (above $0.5~n_{\rm sat}$), and errors of $\lesssim3\%$ at high densities.

\section{Thermal contribution to the energy}
\label{sec:thermal}
We now turn to the thermal energy, which was first defined in eq.~(\ref{eq:Eth}) as
\begin{align*}
\begin{split}
E_{\rm th}(n, Y_p, T) & = E_{\rm nucl, th}(n, Y_p=\sfrac{1}{2},T) \\
	& + E_{\rm sym, th}(n,T) \times (1-2 Y_p)^2 \\ 
	& + E_{\rm lepton, th}(n, Y_p, T) .
\end{split}
\end{align*}

It is useful to further divide the thermal energy into density regimes, over which the matter displays distinct behaviors. At the lowest densities, the contribution from relativistic leptons and photons dominates. At intermediate densities, an ideal-fluid description suffices. However, at high densities, matter can remain partially degenerate even at intermediate-to-high temperatures. In the high-density regime, some of the available energy goes into lifting the degeneracy of the particles rather than adding thermal support and, accordingly, the thermal pressure can dip well below the prediction for an ideal fluid. (See Fig.~\ref{fig:PthNL3} for the markedly different behaviors in thermal pressure across these three regimes.) 

It is, therefore, convenient to write the thermal energy as
\begin{align}
\begin{split}
 E_{\rm th}&(n, Y_p, T)  =  \\
 & \begin{cases}
E_{\rm rel}(n,T),    &\phantom{n_1<} n < n_1 \\
E_{\rm ideal}(T),  &n_1 < n <n_2 \\
E_{\rm th, deg.}(n,Y_p=1/2,T) \\
 \quad+ E_{\rm sym, th}(n, T)  (1-2 Y_p)^2, &\phantom{n_1<} n>n_2
\end{cases}\\
\end{split}
\end{align}
where the relativistic component,
\begin{equation}
E_{\rm rel}(n,T) = \frac{4 \sigma}{c}\frac{f_s}{ n} T^4,
\end{equation}
and the ideal component,
\begin{equation}
E_{\rm ideal}(T) = \frac{3}{2} k_BT
\end{equation}
are given as in eqs.~(\ref{eq:Ethhyb}) and (\ref{eq:fs}). Here, $E_{\rm th, deg.}(n,Y_p=1/2,T)$ is the degenerate thermal energy of symmetric matter, which we introduce below. We note that, because the ideal-fluid and relativistic terms do not depend on the proton fraction, the symmetry-energy correction is only relevant in the degenerate regime. Finally, we define the first transition density, $n_1$, as the density at which the relativistic and ideal-fluid energies are equal. The second transition density, $n_2$, is the density at which the ideal-fluid energy is equal to the degenerate thermal energy, for a given temperature and proton fraction. 

This piecewise expression of the thermal energy is convenient for later calculations of the thermal pressure and the sound speed. However, the discontinuities at the transition densities are artificial and will create problems in numerical simulations, potentially leading to undesired reflections of matter waves at density boundaries. Thus, whenever we actually implement the thermal energy or pressure, we use a smoothed version instead. This smoothed version is of the form
\begin{align}
\begin{split}
\label{eq:addinv}
E_{\rm th}&(n, Y_p, T) \approx E_{\rm rel}(n,T) \\
	&+ \left[ E_{\rm ideal}(T)^{-1} +  E_{\rm th, deg.}(n,Y_p,T)^{-1} \right]^{-1},
\end{split}
\end{align} 
where we have added the latter two terms inversely to ensure that the ideal term dominates at intermediate densities and the degenerate term dominates at the highest densities. The smoothed approximation is also more computationally efficient than the piecewise version, as it does not require the calculation of transition densities, which will vary with the temperature and proton fraction.

In order to calculate the thermal energy in the degenerate regime, we consider the nucleons as a free Fermi gas. In that limit, the leading-order thermal energy of degenerate matter is given by
\begin{align}
\begin{split}
\label{eq:Ethq}
E_{\rm th,~q}^{\rm deg}(n, Y_q, T) &=  a(Y_qn, M^*) \left(\frac{N_q}{N_p + N_n} \right)T^2,\\
	& = a(Y_qn, M^*)  Y_q T^2
\end{split}
\end{align}
for a single-species system of particle $q$.
For simplicity, we have introduced the level-density parameter $a$, which is defined as
\begin{equation}
\label{eq:a}
a(n_q, M^*) \equiv \frac{\pi^2 k_B^2}{2} \frac{ \sqrt{ \left( 3 \pi^2 n_q \right)^{2/3}  (\hbar c)^2 + M^*(n_q)^2 } }{\left( 3 \pi^2 n_q \right)^{2/3} (\hbar c)^2 },
\end{equation}
where $M^*(n_q)$ is the Dirac effective mass of the relevant species at a specific density. \citep[For a complete derivation at next-to-leading order in temperature, see][]{Constantinou2015}.

As an example, the thermal nuclear energy for symmetric matter would be
\begin{align}
\begin{split}
E_{\rm th,~nucl}^{\rm deg}(n, T) &=  \left[ \frac{a(n_p, M^*_{\rm p, SM})N_p + a( n_n, M^*_{\rm n, SM})N_n }{N_p + N_n}\right]  T^2 \\
	&= a(0.5 n, 0.5 M^*_{\rm SM}) T^2, 
\end{split}
\end{align}
where the subscript SM stands for symmetric matter and, in the second line, we have used the fact that $n_n = n_p = 0.5n$ in symmetric matter. We have further made the approximation that the effective masses of neutrons and protons are comparable in symmetric matter and that the average of these two effective masses gives the overall effective mass of symmetric matter, i.e.,  $M^*_{\rm n, SM} \approx M^*_{\rm p, SM} \approx \sfrac{1}{2} M^*_{\rm SM}$.

By likewise defining the thermal energy per baryon for pure neutron matter (PNM), we can calculate the thermal contribution to the symmetry energy, as
\begin{align}
\label{eq:Esymth}
\begin{split}
&E_{\rm sym, th}(n, T) =  \\
& \begin{cases}
 0,   &  n < n_2 \\
 \left[ a(n, M^*_{\rm PNM}) - a(0.5 n, 0.5 M^*_{\rm SM}) \right] T^2, &  n >  n_2,
\end{cases}
\end{split}
\end{align}
 where the low-density limit of $E_{\rm sym, th}$ arises from the fact that both pure neutron matter and symmetric matter behave identically as ideal or relativistic fluids at $n < n_2$. 

In principle, this symmetry energy term extrapolates the thermal energy of symmetric nuclear matter to arbitrary proton fraction. However, we find that including this term has a negligible effect on the results. In particular, making the approximation $E_{\rm th,~nucl}(n, Y_p, T)\approx E_{\rm th,~nucl}(n, Y_p=\sfrac{1}{2}, T)$ introduces an average error of $\lesssim 1\%$  in the total pressure across the density range of interest. We thus neglect the thermal correction to the symmetry energy for the remainder of the paper.

For leptons, the degenerate thermal pressure is even simpler. The effective mass of electrons is approximately constant, due to their small cross-sections of interaction. Hence, $M^*_e \approx m_e$. This allows us to write eq.~(\ref{eq:Ethq}) simply as
\begin{equation}
\label{eq:Ethelect}
E_{\mathrm{th,~}e^-}^{\rm deg}(n, Y_p, T) = a(Y_p n, m_e) Y_p T^2,
\end{equation} 
where we have required that the electron fraction balance the proton fraction in order to satisfy the requirement of charge neutrality and we have used eq.~(\ref{eq:Yp}) to substitute $Y_p$. We note that in the presence of a significant population of positrons, the proton fraction in eq.~(\ref{eq:Ethelect}) should be replaced by the net lepton fraction.

With expressions for the degenerate and ideal fluid thermal terms in hand, we can now write a complete version of eq.~(\ref{eq:Eth}) for $E_{\rm th}$  as follows:
\begin{align}
\begin{split}
\label{eq:Ethfull}
& E_{\rm th}(n, Y_p, T)  =  \\
&
\begin{cases}
4 \sigma f_s  T^4/ (c n),    &\phantom{n_1<} n < n_1 \\
(3/2) k_B T,  &n_1< n <n_2 \\
\left[ a(0.5 n, 0.5 M^*_{\rm SM}) + a(Y_p n, m_e) Y_p   \right] T^2 , &\phantom{n_1<} n>n_2
\end{cases}
\end{split}
\end{align}
where we have neglected the thermal contribution to the symmetry energy, as discussed above.

We thus have a complete expression for the thermal  energy of matter as a function only of the density, temperature, proton fraction, and the effective mass of the nucleons in symmetric matter.

\subsection{$M^*$-approximation}
\label{sec:Mstarapprox}
A full calculation of $E_{\rm th}$ using eq.~(\ref{eq:Ethfull}) requires knowledge of the Dirac effective masses in symmetric matter, and hence the scalar meson interactions and particle potentials of a particular EOS. We instead choose to express the Dirac effective mass with a physically-motivated yet computationally-simple approximation. At low densities, the effective mass must approach the dominant nucleon mass, while at higher densities, $M^*$ must decrease as particle interactions become important. We represent this behavior by introducing a power-law expression,
\begin{equation}
\label{eq:Meff}
M^*(n_q) = \left\{ (m c^2)^{-b} + \left[ mc^2 \left( \frac{ n_q }{n_0 }\right)^{-\alpha} \right]^{-b} \right\}^{-1/b},
\end{equation}
where $m$ is the nucleon mass (which we take to be the neutron mass, $mc^2=939.57$~MeV)\footnote{The EOS in our sample vary in their low-density limit of $M^*$ from 938$-$939.57~MeV. This parameter can easily be adjusted to any low-density value for $M^*$. For simplicity, however, we take it to simply be the neutron mass. We find that this simplification has a negligible effect on our results. } and $n_0$ is the transition density above which $M^*$ starts to decrease. The exponent $b$ determines the sharpness of the transition and $\alpha$ specifies the power-law slope at high densities.
We find that $b=2$ works well to represent the curvature connecting the low- and high-density regimes, and thus fix it to this value in the following analysis, leaving just two free parameters to describe the effective mass, $M^* = M^*(n_0,\alpha)$. 

We fit the effective masses together at $k_BT=1, 10,$ and 47.9~MeV for nine of the EOS in our sample, using a standard least-squares method across the entire density range provided. We exclude the models LS and SLY4-RG here because the effective masses for these EOS are not currently published (but see $\S$\ref{sec:nonRMF} for a separate comparison with these models). The results of these fits are given in Table~\ref{table:Meff} for symmetric matter. For completeness, we also include in Table~\ref{table:Meff} the fits for pure neutron matter, which can be used to calculate $E_{\rm sym,th}(n,T)$ in eq.~(\ref{eq:Esymth}).

\begin{deluxetable}{lcccccc}
\tabletypesize{\footnotesize}
\tablewidth{0.48 \textwidth} 
\tablecaption{\label{table:Meff} Parameters characterizing $M^{*}$, fit together at $k_BT=1, 10,$ and 47.9~Mev, for either pure neutron matter (PNM) or symmetric matter (SM).  }
\tablehead{\\
\colhead{} &
\colhead{PNM ($Y_p$ = 0.01)}  \hspace{-2cm} \vspace{-0.05cm} &
\colhead{} &
\colhead{SM ($Y_p $= 0.5)}  \hspace{-2cm} \vspace{-0.05cm}  &
\colhead{} \\\\
\colhead{ EOS }  &
\colhead{$n_0$ (fm$^{-3}$) }  &
\colhead{$\alpha$ }  &
\colhead{$n_0$ (fm$^{-3}$) }  &
\colhead{$\alpha$ }  
}
\startdata
          TM1 &  0.11 & 0.73   &    0.12 & 0.86    \\
          TMA &  0.11 & 0.65   &    0.13 & 0.77    \\
          NL3 &  0.10 & 0.90   &    0.11 & 1.08    \\
          FSUGold &  0.10 & 0.61   &    0.11 & 0.72    \\
          IUFSU &  0.11 & 0.72   &    0.12 & 0.85    \\
          DD2 &  0.08 & 0.68   &    0.10 & 0.84    \\
        STOS &  0.11 & 0.76   &    0.12 & 0.90    \\
         SFHo &  0.21 & 0.82   &    0.22 & 0.89    \\
         SFHx &  0.16 & 0.77   &    0.17 & 0.88    \\
\hline \\       Range  &  0.08-0.21 & 0.61-0.90     &    0.10-0.22 & 0.72-1.08                         \\ \hline \\
        Mean  &  0.12 & 0.74   &    0.13 & 0.87
\enddata
\tablecomments{We fix $b=2$ and $m=m_n$ in all fits.}
\end{deluxetable}

\begin{figure}[ht]
\centering
\includegraphics[width=0.485\textwidth]{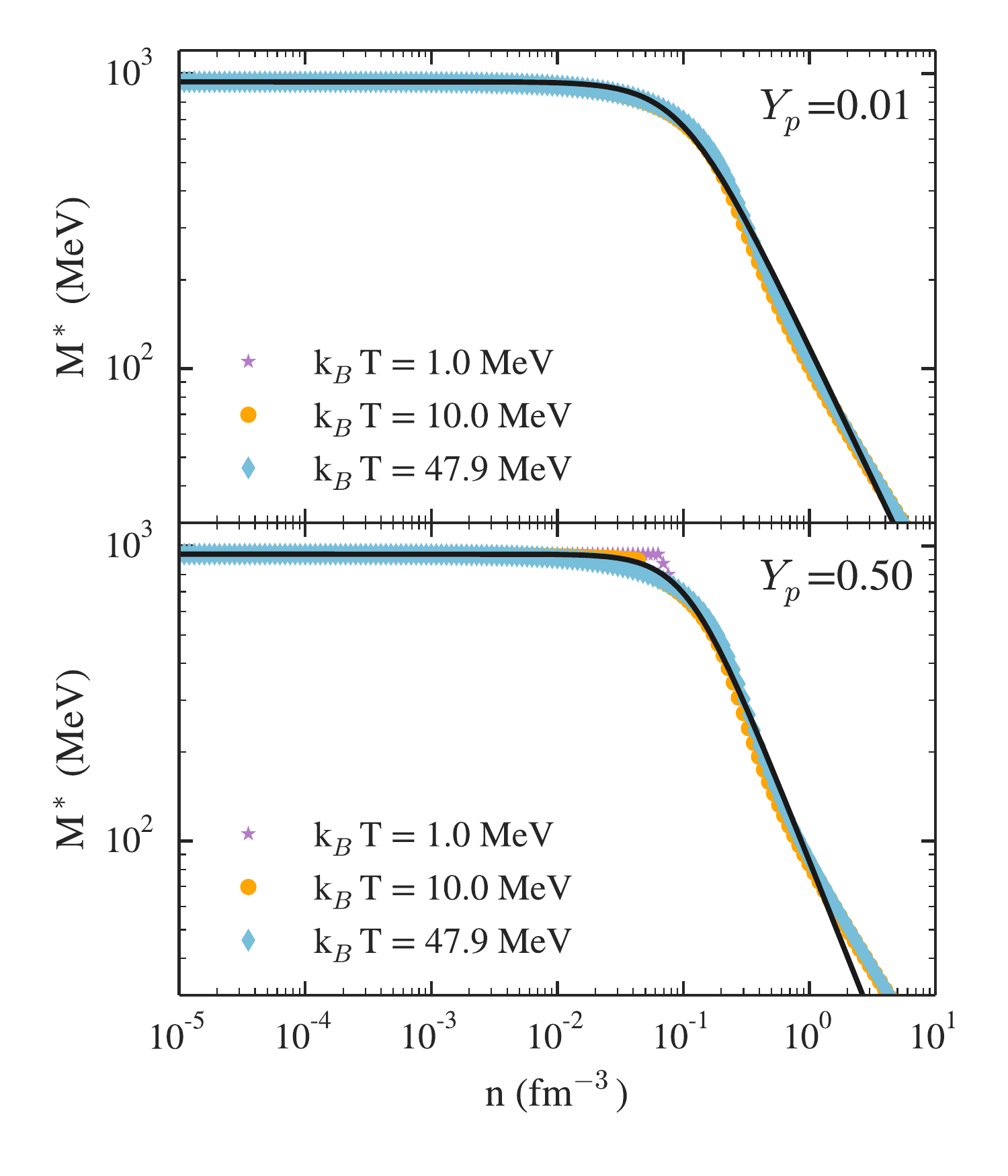}
\caption{\label{fig:Meff} Dirac effective mass as a function of the number density, for NL3 at $k_BT=1, 10$ and 47.9~MeV (in purple, orange, and blue, respectively) for $Y_p$=0.01 (pure neutron matter; top panel) and $Y_p$=0.5 (symmetric nuclear matter; bottom panel). The symbols represent the effective mass predictions for the full version of NL3. The solid black line shows our approximation using eq.~(\ref{eq:Meff}). We find that, with fit parameters $n_0=0.10$~fm$^{-3}$ and $\alpha=0.90$, the $M^*$-approximation accurately reproduces the values predicted by the full EOS for pure neutron matter. For symmetric nuclear matter with $n_0=0.11$~fm$^{-3}$ and $\alpha=1.08$, the $M^*$-approximation again reproduces the values predicted by NL3 reasonably well, up to $\sim10~n_{\rm sat}$. At low temperatures, the discontinuity in the effective mass stems from the Maxwell construction used in the original EOS calculation to represent the phase transition to uniform nuclear matter. At high temperatures, this artifact disappears.}
\end{figure}

We show the performance of the fit for NL3 in Fig.~\ref{fig:Meff}. In this fit, we use the NL3 tables calculated at $k_BT=1, 10$, and 47.9~MeV (shown in purple, orange, and blue, respectively) with a proton fraction of $Y_p$=0.01, to emulate pure neutron matter (top panel), and $Y_p$=0.5, to represent symmetric matter (bottom panel). We show our approximation for $M^*$ as the black solid line. We find that the $M^*$-approximation accurately captures the behavior predicted by the full EOS, with fit parameters $n_0=0.10$~fm$^{-3}$ and $\alpha=0.90$ for $Y_p=0.01$ and fit parameters $n_0=0.11$~fm$^{-3}$ and $\alpha=1.08$ for $Y_p$=0.5. 

As a brief aside, we note a discontinuity in the first derivative of $M^*$ at approximately half the nuclear saturation density for large $Y_p$ and low temperatures (seen most clearly in the purple stars in the bottom panel of Fig.~\ref{fig:Meff}, at $n_{\rm sat}/2 \approx 0.08~$fm$^{-3}$). This discontinuity is an artifact of the treatment of the first-order phase transition to uniform nuclear matter at these densities in the original EOS calculations.

There is an easily understood origin of this artifact. \citet{Lattimer1991}, \citet{Shen1998a}, and \citet{Hempel2010} all use a Maxwell construction to calculate the phase transition at approximately half the nuclear saturation density. At low proton fractions, where matter is approximately made up of a single species, the Maxwell construction works well to represent the phase transition. However, the Maxwell construction is invalid for multi-component species:  When a system has more than one significant component, the Gibbs construction must instead be used \citep{Glendenning1992, Glendenning2000}. Because all EOS that are included in this section use the Maxwell construction, they all suffer from artifacts due to this choice at roughly half the saturation density, where the transition to uniform nuclear matter occurs. 

Correcting these artifacts would require re-calculating all EOS with a different formalism and is beyond the scope of this paper. However, we note that at high temperatures ($k_BT\gtrsim$15~MeV), the non-uniform phase of matter disappears (see discussion around Fig.~5 in \citealt{Shen1998a}). Thus, we can avoid the issue altogether by performing our fit to $M^*$ at only the highest temperatures, when $Y_p$ is large. In practice, we find that whether we fit only the $k_BT=47.9$~MeV curve for $M^*$ or we fit the curves for all the temperatures together, the difference in the resulting parameters is small. We, therefore, choose to perform the fits to three temperatures ($k_BT=1, 10$ and 47.9~MeV) together and use the same method for both low and high proton fractions.

Returning to our discussion of the $M^*$ model, we note that the errors introduced by using our $M^*$-approximation are comparable to those shown in Fig.~\ref{fig:Meff} for the full set of nine EOS in this section. We thus conclude that our $M^*$-approximation reasonably captures the density-dependence of the Dirac effective mass, while greatly simplifying subsequent calculations. 

Moreover,  we find that the range of inferred fit parameters is relatively narrow. In particular, for a wide range of temperatures and EOS, we find that the transition density lies in the range $n_0 \in (0.08,0.22)$~fm$^{-3}$, with an average value of $\sim$0.13~fm$^{-3}$ for both pure neutron matter and symmetric matter. The power-law index characterizing the decay of $M^*$ is similarly well constrained, with $\alpha \in (0.61-0.90)$, with an average value of 0.74 for pure neutron matter; and $\alpha \in (0.72-1.08)$, with a slightly higher average value of 0.87 for symmetric matter. We find only a weak dependence of $n_0$ and $\alpha$ on the temperature, thus suggesting that these parameters could be treated as constants for use in numerical simulations. 

\subsection{Performance of the $M^*$-approximation of thermal effects at fixed $Y_p$}
\label{sec:comparison}

We now turn to a comparison between the $M^*$-approximation of the thermal effects and the nine EOS listed in Table~\ref{table:Meff}. As in $\S$\ref{sec:Esym}, we make the comparison in terms of the pressure, rather than the energy, and use eq.~(\ref{eq:getP}) to convert between the two. The expressions for $P_{\rm th}(n, Y_p, T)$ are given in Box II in $\S$\ref{sec:boxes}. In particular, all results shown here use the smoothed approximation of the thermal pressure, as defined in eq.~(\ref{eq:addinvP}).

In order to focus specifically on the thermal pressure, we calculate the thermal contribution to the pressure from each realistic EOS in our sample by subtracting the cold component at the same $Y_p$.

\begin{figure}[ht]
\centering
\includegraphics[width=0.48\textwidth]{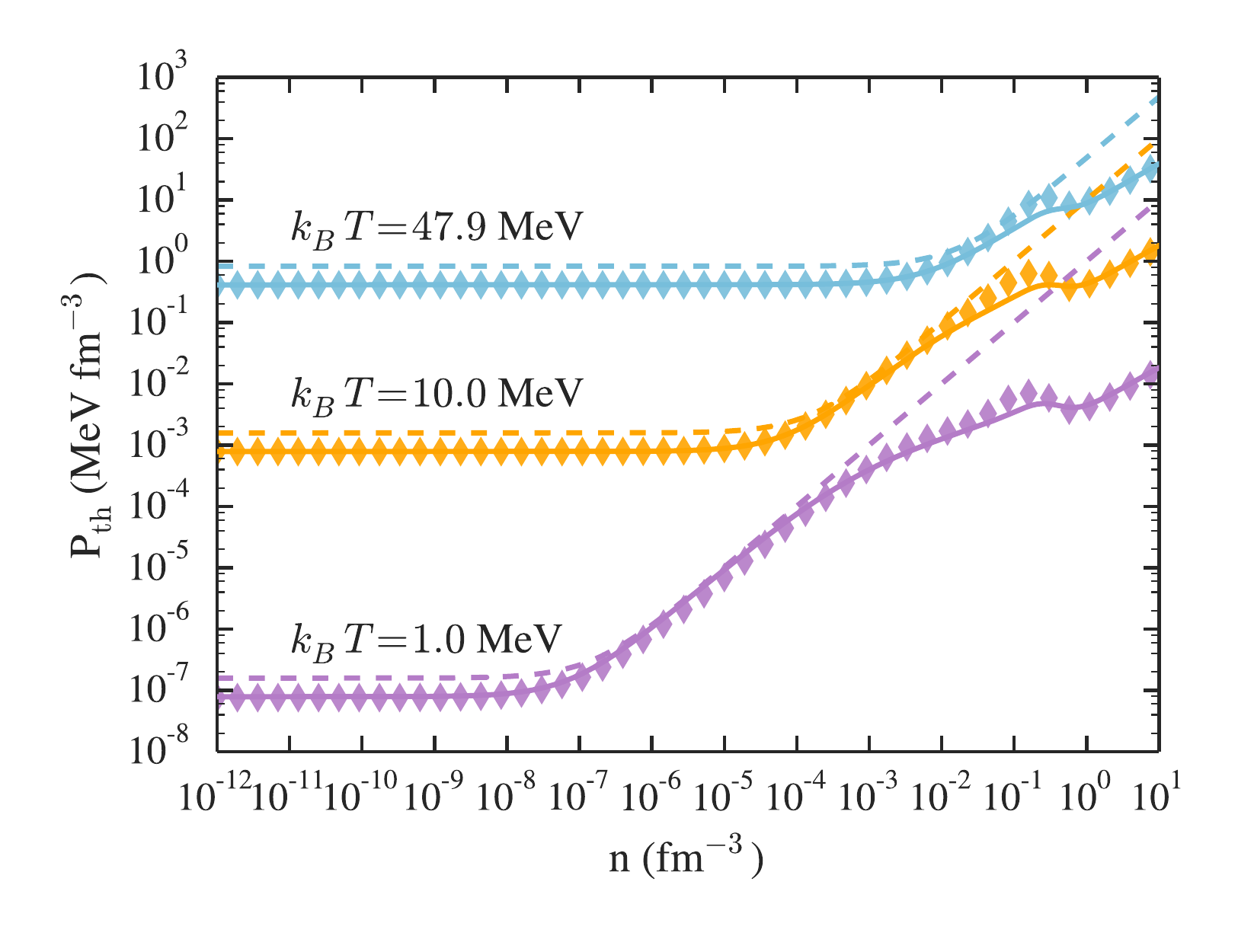}
\caption{\label{fig:PthNL3} Smoothed thermal pressure as a function of density for the EOS NL3 with $Y_p=0.1$. The various colors are calculated at $k_B T$=1~MeV (purple), $k_B T$=10~MeV (orange), and $k_B T$=47.9 MeV (blue). The thermal pressure of the full EOS is shown as the symbols, while the solid lines represent the $M^*$-approximation of $P_{\rm th}$, using the fit parameters for NL3 from Table~\ref{table:Meff} ($n_0 =0.11$~fm$^{-3}, \alpha=1.08$). The dashed lines show the $\Gamma_{\rm th}=1.67$ hybrid approximation at each temperature. We find excellent agreement between the $M^*$-approximation and the full thermal pressure and find that the $M^*$-approximation offers a significant improvement over the hybrid EOS. }
\end{figure}

\begin{figure}[ht]
\centering
\includegraphics[width=0.48\textwidth]{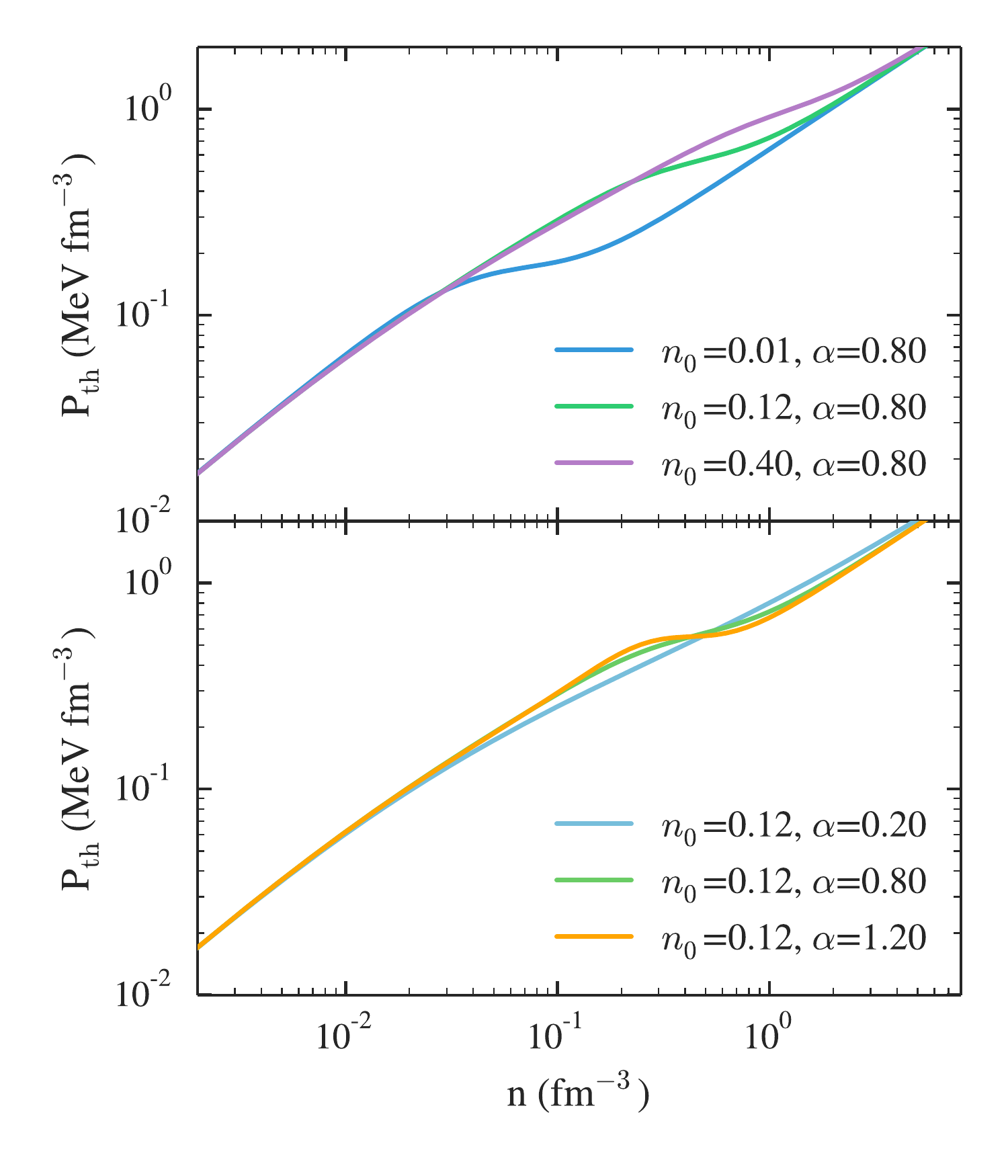}
\caption{\label{fig:varyParams} The $M^*$-approximation of the thermal pressure at $k_BT=10$~MeV and $Y_p=0.5$, with intentionally extreme choices of the parameter values. The top panel shows the effect of varying $n_0$ for a fixed value of $\alpha=0.8$; the bottom panel shows the effect of varying $\alpha$ for fixed $n_0$=0.12~fm$^{-3}$.   }
\end{figure}
\begin{figure*}[ht]
\centering
\includegraphics[width= \textwidth]{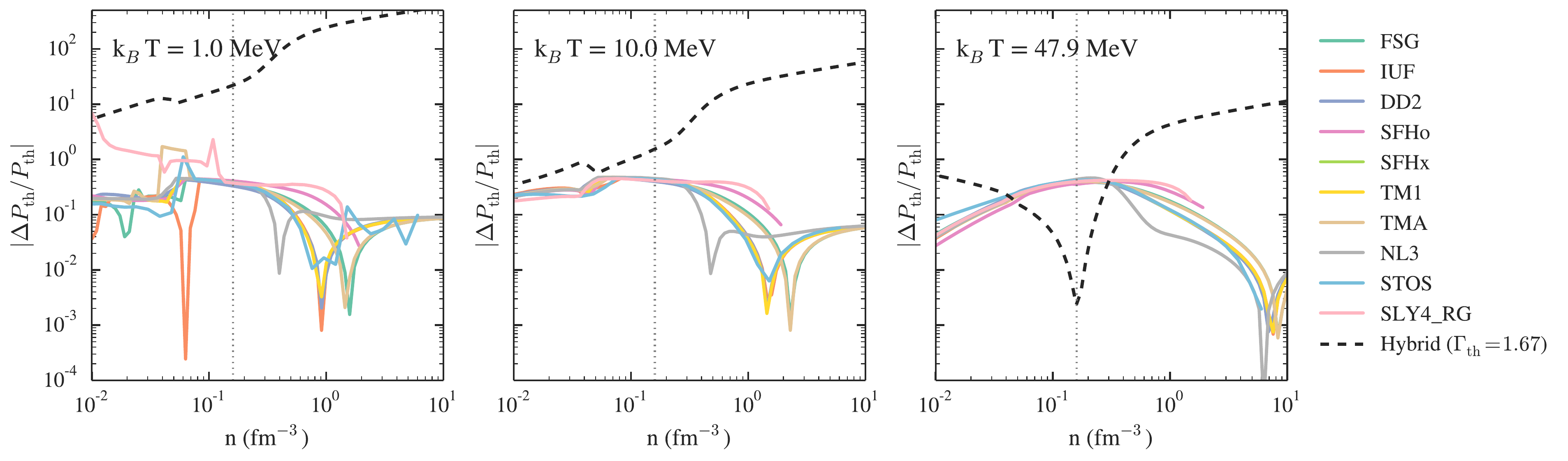}
\caption{\label{fig:resids} Residuals between the smoothed $M^*$-approximation of the thermal pressure and the full results calculated for each EOS listed in Table~\ref{table:Meff}. From left to right, the panels are at $k_BT=1, 10$ and 47.9~MeV; all three panels are for $Y_p=0.1$. The various colors represent the different EOS. For comparison, we also include the residuals between the full EOS NL3 and the ideal-fluid approximation ($\Gamma_{\rm th}=1.67$) as the black dashed line. The vertical dotted line marks $n_{\rm sat}$. Our M$^*$-approximation of $P_{\rm th}$ produces residuals that are up to three orders of magnitude smaller than the ideal-fluid approximation.  }
\end{figure*}

In general, we find excellent agreement between the $M^*$-approximation and the thermal pressures calculated from the full EOS. We show an example in Fig.~\ref{fig:PthNL3} for NL3. We find that our approximation of $P_{\rm th}$ closely recreates the full calculation for NL3 for nearly all densities and temperatures explored here. For comparison, we also include in Fig.~\ref{fig:PthNL3} the hybrid approximation with $\Gamma_{\rm th}=1.67$ as dashed lines.\footnote{We choose the relatively low value of $\Gamma_{\rm th}=1.67$ in order to minimize the residuals of the hybrid model. This value of $\Gamma_{\rm th}$ ensures the hybrid EOS matches an ideal fluid at intermediate densities. Larger values, as are more commonly used in numerical simulations, would cause the hybrid $P_{\rm th}$ to overestimate even the ideal regime.}  The full thermal pressure agrees with the hybrid approximation only at intermediate densities. At the lowest densities, this value of $\Gamma_{\rm th}$ overestimates the contribution from relativistic species.  At higher densities that are relevant for forming and merging neutron stars, particle interactions become important and the ideal-fluid approximation grossly overestimates the thermal pressure, remaining several orders of magnitude above the true thermal pressure.

In order to gain an intuitive understanding of the behavior of $P_{\rm th}$, we also explore an extreme range of the $M^*$ parameters. Specifically, in Fig.~\ref{fig:varyParams}, we zoom in on $P_{\rm th}$ at $k_BT=10$~MeV and $Y_p=0.5$ and show the effect of varying the parameters $n_0$ and $\alpha$ for symmetric matter. We intentionally take extreme values for the parameters, well beyond the ranges found in Table~\ref{table:Meff}, in order to emphasize that the variations between more realistic parameter choices will be small. Even for these unreasonable choices of values for $n_0$ and $\alpha$, we find that $P_{\rm th}$ approximates the full thermal pressure reasonably well and, in all cases, better than the ideal fluid approximation. Analyzing the specific dependences more closely, we see in Fig.~\ref{fig:varyParams} that the parameter $n_0$ controls the density at which the rise in the thermal pressure starts to slow. This corresponds to the density at which particle interactions become significant and degenerate thermal effects can no longer be ignored. The parameter $\alpha$, which controls the power-law slope of $M^*$, directly controls the height of the dip in $P_{\rm th}$. This makes intuitive sense:  if particle interactions are stronger, $M^*$ decreases more rapidly, $\alpha$ will be larger, and the thermal pressure will deviate even more drastically from the ideal-fluid approximation as part of the free energy is taken up by those interactions.

Finally, we compare the $M^*$-approximation of the thermal pressure against the full sample of EOS listed in Table~\ref{table:Meff}. We show the corresponding residuals at three temperatures in Fig.~\ref{fig:resids} and find that the residuals are typically $\lesssim30\%$ at densities above $0.5~n_{\rm sat}$. For comparison, Fig.~\ref{fig:resids} also shows a sample set of residuals between the full thermal pressure from NL3 and the hybrid approximation ($\Gamma_{\rm th}=1.67$) as the black dashed line. We find that the $M^*$-approximation  produces residuals that are up to three orders of magnitude smaller than the ideal-fluid approximation used in hybrid EOS, with only two additional parameters that are easy to specify.

\subsection{$M^*$-approximation for non-RMF models}
\label{sec:nonRMF}
We have so far only calculated the thermal pressures using the sub-sample of EOS for which there exist published tables of the effective masses. While this allowed us to directly test the performance of the $M^*$-approximation, this set of EOS happens to also be calculated exclusively with RMF models. In this section, we compare the $M^*$-approximation to the LS and SLY4-RG models, which are calculated using non-relativistic Skyrme energy functionals (see $\S$~\ref{sec:overview}). We also include here the two-loop exchange model of \cite{Zhang2016}, which is an extension of mean field theory. We note that the pressures of the \cite{Zhang2016} EOS are reported only at $Y_p=0$ and 0.5, which is why this EOS is not included in our full sample. As a result of these and other limitations in the publicly-available values for this EOS, all comparisons in this section are made at $Y_p=0.5$ and $T=20$~MeV. We also fix $n_0$ and $\alpha$ to the mean values for symmetric matter from Table~\ref{table:Meff} for all three EOS.

\begin{figure}[ht]
\centering
\includegraphics[width=0.45\textwidth]{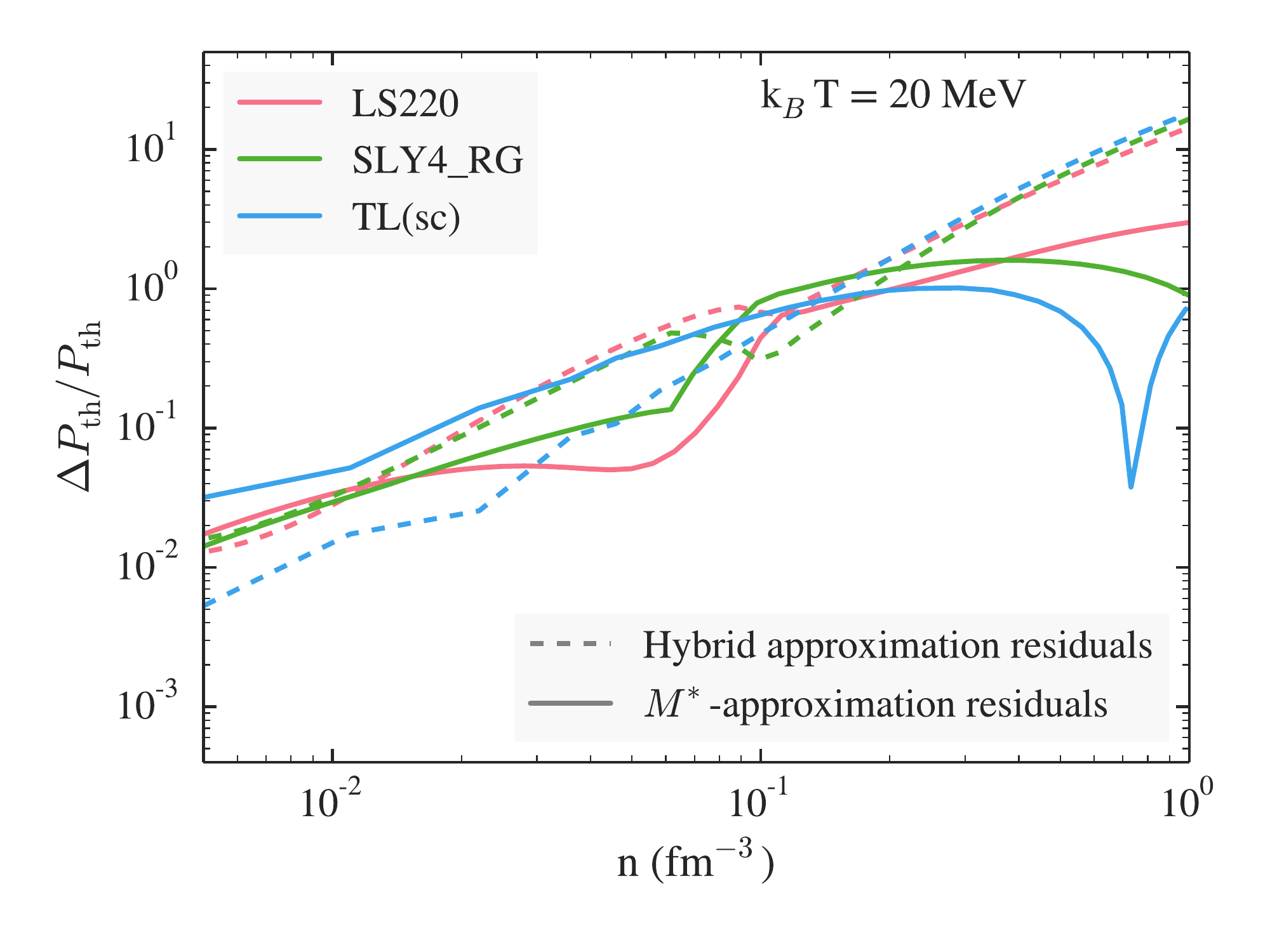}
\caption{\label{fig:nonRMF} Residuals between the smoothed $M^*$-approximation of the thermal pressure and the true EOS at $Y_p$=0.5 and $T=20$~MeV for three non-RMF models. For $n_0$ and $\alpha$, we use the mean fit values for symmetric matter from Table~\ref{table:Meff}. The dashed lines show the corresponding residuals between the true EOS and the hybrid approximation using $\Gamma_{\rm th}$=1.67, at the same proton fraction and temperature. The three EOS shown are LS (pink), SLY4-RG (green), and the two-loop model of \cite{Zhang2016} (``TL(sc)", blue). We find that, while the $M^*$-approximation produces slightly larger residuals for these EOS than for the RMF models, it nevertheless offers a significant improvement over the hybrid approximation at high densities. }
\end{figure}

Figure~\ref{fig:nonRMF} shows the residuals between the $M^*$-approximation of the thermal pressure and the true EOS for these three models. For comparison, this figure also shows the corresponding residuals between the hybrid approximation and the true EOS (dashed lines). In general, we find that the $M^*$-approximation of the thermal pressure results in larger residuals for these EOS compared to the RMF models, but that it still offers a significant improvement over the hybrid approximation at densities above $\sim n_{\rm sat}$.

We also compared the residuals at $T=50$~MeV and found that the $M^*$-approximation performed comparably to the hybrid approximation at this temperature. In fact, for densities between $n_{\rm sat}$ and $0.7$~fm$^{-3}$, the hybrid approximation produces slightly smaller residuals in the thermal pressure for these non-RMF models. In this regime, the hybrid approximation tends to over-estimate the thermal pressure for these models, while the $M^*$-approximation tends to under-estimate $P_{\rm th}$ by a similar degree. However, even in this case, the $M^*$-approximation still offers an appreciable improvement over the hybrid approximation at the highest densities, above $\sim0.7$~fm$^{-3}$.

\section{Putting it all together}
\label{sec:boxes}
We now summarize the equations and approximations that we have developed so far to represent the total energy per particle in Box I. 

\begin{widetext}
\setlength{\fboxrule}{2pt}
\setlength{\fboxsep}{6pt}
\fbox{\parbox{ 0.9\textwidth}{
\textbf{ Box I: Total Energy Expressions for Finite-Temperature Dense Gas.} \\\\
The energy per particle of \npe~matter is given by
\begin{align*}
E(n, Y_p, T)  & = \big( \text{Cold EOS in $\beta$-equilibrium}\big) 
		+ 3 K  \left( Y_p^{4/3} - Y_{p,\beta}^{4/3} \right) n^{1/3} \\
	& + E_{\rm sym}(n, T=0)\left[ (1-2 Y_p)^2 - (1-2 Y_{p,\beta})^2\right] 
 	 +	
\begin{cases}
4 \sigma f_s  T^4/ (c n),    &\phantom{n_1<} n < n_1 \\
(3/2) k_B T,  &n_1< n <n_2 \\
\left[ a(0.5 n, 0.5 M^*_{\rm SM}) + a(Y_p n, m_e) Y_p   \right] T^2 , &\phantom{n_1<} n>n_2,
\end{cases}
\end{align*}
\\
where the symmetry energy is approximated as
\\
\begin{equation*}
E_{\rm sym}(n, T=0) =  \eta E_{\rm sym}^{\rm kin}(n) + \left[ S_0 - \eta E_{\rm sym}^{\rm kin}(n_{\rm sat})\right]\left(\frac{n}{n_{\rm sat}}\right)^\gamma,
\end{equation*}
\begin{equation*}
E_{\rm sym}^{\rm kin}(n) = \frac{3 }{5}  \left( 1 -2^{1/3} \right) E_f(n), 
\end{equation*}
\begin{equation*}
\eta = \frac{5}{9} \left[ \frac{ L-3 S_0 \gamma}{\left(1-2^{1/3}\right)\left(2/3 - \gamma \right) E_f(n_{\rm sat})} \right],
\end{equation*}
\\
and the terms of the $M^*$-approximation are given by
\\
\begin{equation*}
a(n_q, M_q^*) \equiv \frac{\pi^2 k_B^2}{2} \frac{ \sqrt{ M_q^*(n_q)^2 + (3 \pi^2 n_q)^{2/3}(\hbar c)^2 } }{(3 \pi^2 n_q)^{2/3}(\hbar c)^2 }
\quad \text{and} \quad
M^*(n_q) = \left\{ (m c^2)^{-b} + \left[ mc^2 \left( \frac{ n_q }{n_0 }\right)^{-\alpha} \right]^{-b} \right\}^{-1/b}.
\end{equation*}
\begin{itemize}
\item The parameters $S_0$, $L$, and $\gamma \in (0.2-1.2)$ are freely specified; this will uniquely specify $Y_{p,\beta}$. 
\item Alternatively, $S_0$, $L$, and $Y_{p,\beta}$ may be specified and the proton fraction may be fit for $\gamma$. We provide fits to $\gamma$ for eleven EOS in Table~\ref{table:gamma}.
\item We find that for $M^*_{\rm SM}$, $n_0 \sim 0.13$~fm$^{-3}$ and $\alpha \sim 0.9$ provide reasonable fits to most EOS.
\end{itemize}
}}
\end{widetext}

Using the expressions for the energy from Box I., we can derive the pressure via the standard thermodynamic relations of eq.~(\ref{eq:getP}), where the derivatives are evaluated at constant $Y_p$, $Y_{p,\beta}$, and $S$. The total entropy of the relativistic, ideal-fluid, and degenerate terms is given by
\begin{align}
\begin{split}
S(n,N_p, N_n,N_e, T) =
 \begin{cases}
S_{\rm rel},    &\phantom{n_1<} n < n_1 \\
S_{\rm ideal} ,  &n_1< n <n_2 \\
S_{\rm deg} , &\phantom{n_1<} n>n_2,
\end{cases}
\end{split}
\end{align}
where $n_1$ and $n_2$ are the thermal energy transition densities, as defined in $\S$\ref{sec:thermal}.

The entropy of a gas of relativistic leptons and photons is given by
\begin{equation}
S_{\rm rel} =  \frac{16 \sigma f_s}{3 c} \left( \frac{N_p + N_n}{n}\right) T^3.
\end{equation}

The entropy of a monatomic ideal fluid is given by the Sackur-Tetrode equation,
\begin{multline}
S_{\rm ideal} =  \left( N_p + N_n + N_e \right) k_B \\
\times \left\{ \ln\left[ \left( \frac{N_p+N_n}{N_p + N_n + N_e} \right) n^{-1} \left(\frac{ m k_B T}{ 2\pi \hbar^2}\right)^{3/2} \right] + \frac{5}{2} \right\}.
\end{multline}

Finally, the entropy of a degenerate Fermi gas in our framework is given by
\begin{equation}
\label{eq:S}
S_q = 2 a_q N_q T
\end{equation}
for a particle $q$, so that the total entropy for the degenerate terms is
\begin{equation}
S_{\rm deg} =2 \left\{ a(0.5 n, 0.5 M^*_{\rm SM})[N_n + N_p] + a(Y_p n, m_e) N_e \right\} T.
\end{equation}

We summarize the resulting pressure equations in Box II.

\begin{widetext}
\setlength{\fboxrule}{2pt}
\setlength{\fboxsep}{6pt}
\fbox{\parbox{ 0.9\textwidth}{
\textbf{ Box II: Pressure Expressions for Finite-Temperature Dense Gas.} \\\\
The pressure of \npe~matter is given by
\begin{align*}
P(n, Y_p, T)&  = \big( \text{Cold EOS in $\beta$-equilibrium}\big) 
	+ K  \left( Y_p^{4/3} - Y_{p,\beta}^{4/3} \right) n^{4/3} \\
	& + P_{\rm sym}(n, T=0)\left[ (1-2 Y_p)^2 - (1-2 Y_{p,\beta})^2 \right] 
 +
 \begin{cases}
4 \sigma f_s  T^4/ (3 c),    &\phantom{n_1<} n < n_1 \\
n k_B T,  &n_1< n <n_2 \\
-\left[ \frac{\partial a(0.5 n, 0.5 M^*_{\rm SM})}{\partial n} + \frac{\partial  a(Y_p n, m_e)}{\partial n} Y_p    \right] n^2 T^2 , &\phantom{n_1<} n>n_2,
\end{cases}
\end{align*}
\\
where $n_1$ and $n_2$ are the thermal energy transition densities for a particular temperature and proton fraction.

The symmetry pressure, corresponding to our model of the symmetry energy, is 
\\
\begin{align*}
P_{\rm sym}(n, T=0) & = \frac{2\eta}{3}  n E_{\rm sym}^{\rm kin}(n) + 
	\left[ S_0 - \eta E_{\rm sym}^{\rm kin}(n_{\rm sat})\right] \left( \frac{n}{n_{\rm sat}}\right)^{\gamma} \gamma n  . 
\end{align*}
\\
The full analytic expression for $Y_{p,\beta}$ is given in eq.~(\ref{eq:YpBeta}) and derived in Appendix A.
\\
The $M^*$-approximation derivatives are given by
\\
\begin{multline*}
\frac{\partial a(n_q, M^*)}{\partial n}\biggr\rvert_{Y_q} = 
-\frac{2 a(n_q, M^*)}{3 n}  
\left\{ 1 -\frac{1}{2}\left[ \frac{M^*(n_q)^2}{ M^*(n_q)^2+(3\pi^2 n_q)^{2/3}(\hbar c)^2} \right] \left( \frac{(3\pi^2 n_q)^{2/3}(\hbar c)^2}{M^*(n_q)^2}+ 3 \frac{\partial \ln [ M^*(n_q)]}{\partial \ln n}\biggr\rvert_{Y_q}  \right) \right\},
\end{multline*}
\quad and
\begin{equation*}
\frac{\partial \ln[ M^*(n_q)]}{\partial \ln{n}} \biggr\rvert_{Y_q} = -  \alpha \left[ 1- \left(\frac{M^*(n_q)}{Y_q mc^2}\right)^2 \right],
\end{equation*}
where, for symmetric matter, we replace $M^*({n_q})\rightarrow 0.5 M^*_{\rm SM}(0.5 n)$ and for the electrons, $M^*({n_q}) \rightarrow m_e$.
\begin{itemize}
\item As in Box I., there are five free parameters:  $S_0, L, \gamma, n_0, \alpha$. 
\item A user may freely specify $S_0$, $L$, and $Y_{p,\beta}$ and fit for $\gamma$. Alternatively, a user may specify  $S_0$, $L$, and $\gamma$, which will uniquely specify $Y_{p,\beta}$.
\item We provide fits for $\gamma, n_0,$ and $\alpha$ for the EOS in our sample in Tables~\ref{table:gamma} and \ref{table:Meff}.
\end{itemize}
}}
\end{widetext}

\begin{figure*}[ht]
\centering
\includegraphics[width=0.95\textwidth]{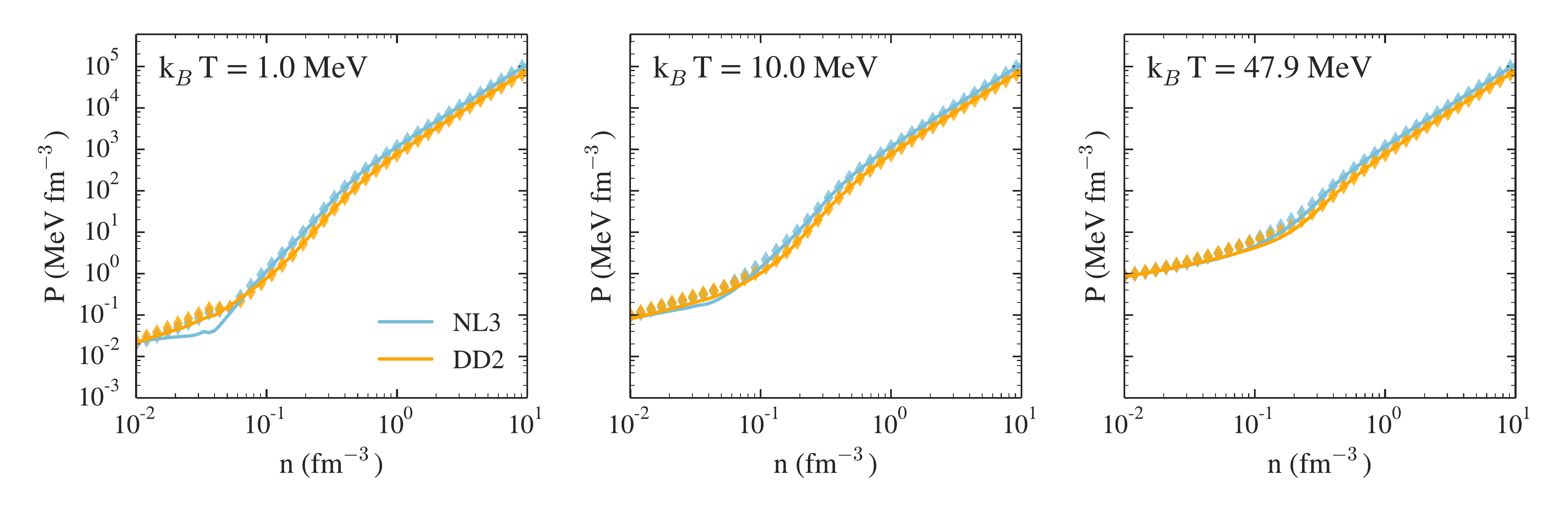}
\caption{\label{fig:pFull} Our approximation of $P$ and the EOS pressures predicted by NL3 and DD2 (in blue and orange, respectively). The EOS predictions are shown as the diamonds, while our model is shown as the solid lines. The three panels are at $Y_p=0.1$ and $k_BT=1, 10,$ or 47.9~MeV (from left to right). We find that our approximation is able to closely recreate the pressures predicted by NL3 and DD2 at densities above $n_{\rm sat}$ for all temperatures. }
\end{figure*}

\begin{figure*}[ht]
\centering
\includegraphics[width=0.98\textwidth]{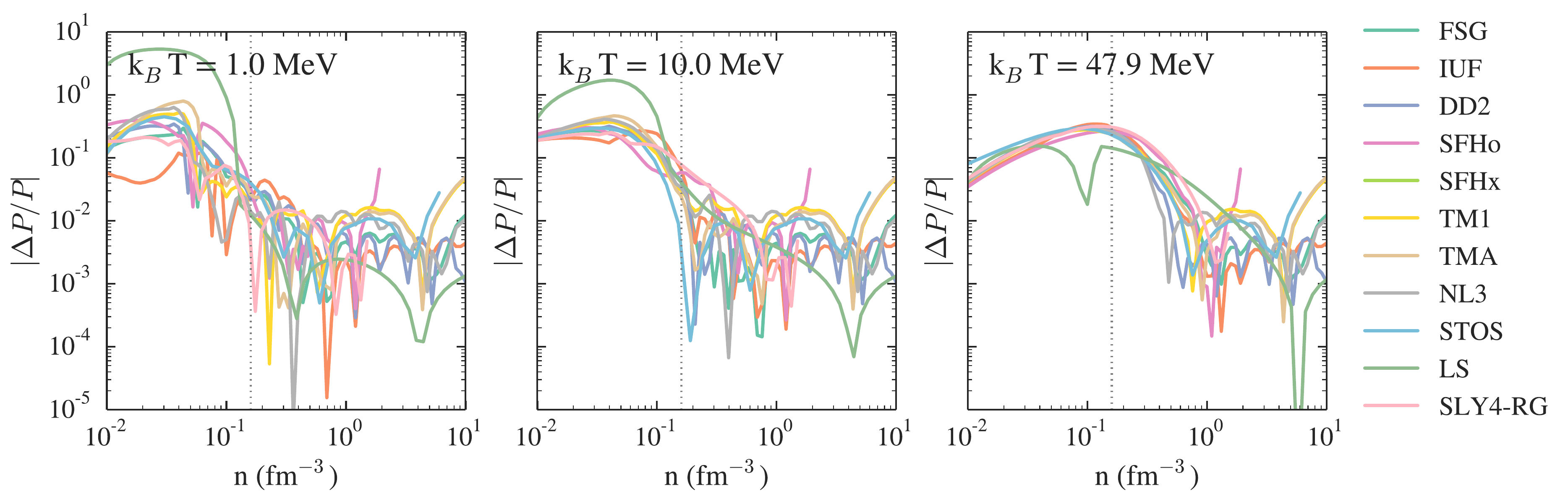}
\caption{\label{fig:residsFull} Residuals between our approximation of $P$ and the EOS pressures predicted by the eleven EOS in our sample. The three panels are at $Y_p=0.1$ and $k_B T=1, 10,$ or 47.9~MeV (from left to right). The vertical dotted line marks $n_{\rm sat}$. }
\end{figure*}

The piecewise definitions of the thermal energy and pressure are mathematically convenient, but the sharp transitions are themselves unphysical, as discussed in $\S$\ref{sec:thermal}. We, therefore, instead implement the thermal pressure using a smoothed approximation of the form
\begin{equation}
\label{eq:addinvP}
P_{\rm th}(n, Y_p, T) \approx P_{\rm rel} + \left(  P_{\rm ideal}^{-1} +  P_{\rm deg}^{-1} \right)^{-1}.
\end{equation}
This smoothed approximation of the thermal pressure is used for the figures throughout this paper. We note that we use this separate smoothing for both the thermal pressure and the thermal energy (as in eq.~\ref{eq:addinv}) in order to keep the problem tractable. However, this is not mathematically exact since, formally, the energy is the proper thermodynamic function and the pressure should, ideally, be derived from the smoothed energy. Nevertheless, the errors introduced by the separate smoothing approximations will be limited to the regions close to the transition points. Physically, the mismatch between the approximate thermal energy and pressure will correspond to a small error in the sound speed in these regions, which we neglect for the present purposes. 

Finally, we note that our model allows significant freedom in creating a new finite-temperature EOS. We have provided a set of parameters that correspond to physically-motivated EOS, but if one wishes to vary these parameters significantly, it will be useful to check that the resulting EOS is still physical. One requirement of a realistic EOS is that the sound speed remain sub-luminal at all densities and temperatures of interest. For this reason, we include in Appendix B a calculation of the sound speed for astrophysical merger scenarios.

\section{Complete model:  Comparison of realistic EOS at arbitrary $Y_p$ and $T$}
\label{sec:complete}
In $\S$\ref{sec:Esym}, we found that our model is able to extrapolate from $\beta$-equilibrium to an arbitrary proton fraction with resulting errors of $\lesssim10\%$ at densities above 0.5~$n_{\rm sat}$. Similarly, in $\S$\ref{sec:comparison}, we showed that the $M^*$-approximation is able to reproduce the thermal pressure of realistic EOS, at fixed $Y_p$, to within $\sim30\%$ for a variety of EOS based on RMF theory. In this section, we quantify the performance of our complete model:  starting with a cold EOS in $\beta$-equilibrium, and extrapolating to arbitrary temperature and proton fraction.

Figure~\ref{fig:pFull} shows an example of a complete model for NL3 and DD2 at three different temperatures. For our approximation, we start with the relevant cold EOS in $\beta$-equilibrium and add the corrections outlined in Box~II, to extrapolate the pressure to $Y_p=0.1$ and the three indicated temperatures. We take the values for $n_0, \alpha$, and $\gamma$ listed in Tables~\ref{table:gamma} and \ref{table:Meff} for each EOS. We show the results as the solid lines in Fig.~\ref{fig:pFull}, while the predictions of the full EOS are shown as the diamonds. We find close agreement between our approximation and the full pressures predicted by NL3 and DD2, especially at densities above $\sim0.5~n_{\rm sat}$. 

Figure~\ref{fig:residsFull} shows the corresponding residuals between our approximation and the full EOS for NL3 and DD2, as well as the rest of our sample of EOS. For each EOS in this figure, we use the values for $n_0, \alpha$ and $\gamma$ listed in Tables~\ref{table:gamma} and \ref{table:Meff}, where possible. For LS and SLY4-RG, for which we do not have fit values for $n_0$ and $\alpha$, we use the average parameter values for symmetric matter in Table~\ref{table:Meff}. We find that our approximation works comparably well to recreate any of the EOS in our sample. Moreover, we find that for $n \gtrsim n_{\rm sat}$, the residuals are $\lesssim 20\%$ at all three temperatures.

For all the EOS in our sample, the error introduced by our model increases in the vicinity of $\sim 0.5 n_{\rm sat}$. This is a result of the break-down in the $E_{\rm sym}$ approximation at low densities. Our derivation of $E_{\rm sym}$ in $\S$\ref{sec:Esym} assumed uniform \npe~matter, but at densities below $\sim0.5~n_{\rm sat}$,  the matter becomes inhomogeneous. Nevertheless, with the exception of LS, the errors at these densities are still typically $\lesssim50$\%.

We have thus verified that our model is able to recreate realistic EOS at relevant densities, with a simple set of parameters. The implications of this result are two-fold. First, this approximation can be used in lieu of more complicated calculations, to analytically represent the EOS that are commonly used in the literature with reasonable accuracy. Second, it implies that our approximation can be reliably used to create new finite-temperature EOS for \npe~matter that probe different physics through the choice of $n_0, \alpha, \gamma, S_0,$ and $L$. Our model allows further freedom in creating a new finite-temperature EOS through the choice of the cold, $\beta$-equilibrium EOS. We thus find that this model can span a broad range of possible physics, with parameters that are directly tied to the underlying physics and that can be integrated with minimal computational cost to a large array of numerical calculations.

\section{Conclusions}

In this paper, we have developed a general framework for calculating the pressure of neutron-star matter at arbitrary proton fraction and finite temperature. Our model is designed so that the corrections we have developed here can be added to any cold \npe~EOS in neutrinoless $\beta$-equilibrium. The model is based on a set of five physically-motivated parameters: $S_0, L, \gamma, n_0,$ and $\alpha$. The first three, $S_0$, $L$, and $\gamma$ characterize the symmetry energy and can be chosen to match a particular EOS or set of priors from laboratory experiments. The parameters $n_0$ and $\alpha$ are introduced through our $M^*$-approximation, where $n_0$ represents the density at which particle interactions become important and $\alpha$ characterizes the strength of those interactions. We find that the effective masses of nine realistic EOS can be well characterized by our $M^*$-approximation with a relatively narrow range of these parameters, with average values of $n_0\sim 0.13$~fm$^{-3}$ and $\alpha \sim 0.9$. 

The complete model is able to extrapolate from cold matter in $\beta$-equilibrium to arbitrary proton fraction and temperature. We find that our model is able to recreate a sample of eleven realistic EOS with resulting errors of $\lesssim20\%$ at a variety of temperatures and proton fractions, above $n_{\rm sat}$.  In particular, by including the effects of degenerate matter, our $M^*$-approximation reproduces the thermal pressure of realistic EOS with residuals that are several orders of magnitude smaller than the hybrid EOS that are commonly used in the literature.

In addition to providing a $1-3$ orders-of-magnitude improvement over the ideal-fluid approximation of the thermal pressure, this model also includes the effects of changing the proton fraction, which is particularly relevant in simulating the formation and cooling of proto-neutron stars.

The complete model can thus be used to accurately recreate the realistic EOS that are currently in use in the literature with a set of simple, analytic functions. Furthermore, the model can be used to calculate new finite-temperature EOS that span a wide range of underlying physics, following one of two possible paths. One possibility is to choose a physically-motivated cold EOS, which will provide predictions for the $\beta$-equilibrium proton fraction and symmetry energy parameters. These can then be used to fit for the free parameter $\gamma$, and then used to extrapolate to an arbitrary proton fraction. Alternatively, one can use a cold, parametric EOS that does not specify the microphysics. In this case, there is freedom to choose the symmetry energy parameters to probe entirely new physics. In either case, one can freely choose the interaction parameters to control the relative importance of thermal effects. All together, these possibilities will allow a new and wide range of physics to be robustly probed in studies of dynamical neutron star phenomena.

{\em{Acknowledgements.\/}} We thank Vasileios Paschalidis for useful discussions and comments on this work. CR is supported by the NSF Graduate Research Fellowship Program Grant DGE-1143953. FO and DP acknowledge support from NASA grant NNX16AC56G.

\bibliography{thermalPbib}
\bibliographystyle{apj}

\FloatBarrier
\appendix

\section{A. Relationship between the $\beta$-equilibrium proton fraction and the symmetry energy for \npe~matter }
\label{sec:causality}
In this appendix, we derive the relationship shown in eq.~(\ref{eq:YpBeta}), which asserts that the cold $\beta$-equilibrium proton fraction is uniquely specified by the symmetry energy.

At zero-temperature for \npe~matter, the total energy per baryon is given by
\begin{equation}
E_{\rm tot}(n, Y_p) = E_n(n, Y_n) + E_p(n, Y_p) + E_e(n, Y_e),
\end{equation}
where $E_n$ is the energy per baryon of neutrons, $E_p$ is the energy per baryon of protons, and $E_e$ the energy per baryon of electrons. Here, we also introduce the neutron fraction, $Y_n \equiv (1-Y_p)$, and the electron fraction, $Y_e = Y_p$, where the latter equality holds in a charge-neutral system.

In order to find the minimum of the total energy, we differentiate with respect to $Y_p$ and get
\begin{equation}
\label{eq:E1}
\frac{ \partial E_{\rm tot}(n, Y_p) }{ \partial Y_p}  = 
 \frac{\partial E_n}{\partial Y_n} \frac{\partial Y_n}{\partial Y_p} +
 \frac{\partial E_p}{\partial Y_p}  +
 \frac{\partial E_e}{\partial Y_e} \frac{\partial Y_e}{\partial Y_p},
\end{equation}
where all the partial derivatives here and throughout this appendix are evaluated at constant entropy and baryon density and we have suppressed the notation for clarity. By substituting in the chemical potential of a species $i$, given by $\mu_i \equiv  \partial E_i/ \partial Y_i |_{S,n}$, eq.~(\ref{eq:E1}) simplifies to
\begin{equation}
\label{eq:E1a}
\frac{ \partial E_{\rm tot}(n, Y_p) }{ \partial Y_p} = -\mu_n + \mu_p + \mu_e,
\end{equation}
which is zero in $\beta$-equilibrium. 

Alternatively, we can write the total energy as an expansion about nuclear symmetric matter with electrons added, i.e.,
\begin{equation}
E_{\rm tot}(n, Y_p) = E_{\rm nucl}(n, 1/2) + E_{\rm sym}(n) (1-2 Y_p)^2 + E_e(n, Y_e).
\end{equation}
This results in
\begin{equation}
\label{eq:E2}
\frac{ \partial E_{\rm tot}(n, Y_p) }{ \partial Y_p} = -4 (1-2 Y_p)  E_{\rm sym}(n) +  \frac{\partial E_e}{\partial Y_e} \frac{\partial Y_e}{\partial Y_p}.
\end{equation}
By charge neutrality and the definition of the chemical potential, the second term is simply $\mu_e$. 
Combining eqs.~(\ref{eq:E1a})~and~(\ref{eq:E2}) in $\beta$-equilibrium gives
\begin{equation}
\label{eq:E3}
\mu_e = 4(1-2 Y_{p,\beta}) E_{\rm sym}(n).
\end{equation}
For relativistic electrons, 
\begin{equation}
\mu_e = \sqrt{p_{f}^2c^2 + m_e^2 c^4} \approx p_{f}c,
\end{equation}
where $p_{f}c = (3 \pi^2 Y_e n)^{1/3} \hbar c$ is the Fermi momentum of the electrons. Combining this expression for $\mu$ with eq.~({\ref{eq:E3}) yields
\begin{equation}
(3 \pi^2 Y_{p,\beta} n)^{1/3} \hbar c = 4 (1-2 Y_{p,\beta})E_{\rm sym}(n),
\end{equation}
or, rearranged to match eq.~(\ref{eq:YpBeta}),
\begin{equation}
\frac{Y_{p,\beta}}{(1-2 Y_{p,\beta})^3} = \frac{64 E_{\rm sym}(n)^3}{ 3 \pi^2 n (\hbar c)^3}.
\end{equation}
Solved for $Y_{p,\beta}$, this gives
\begin{equation}
Y_{p,\beta} = \frac{1}{2} + \frac{(2 \pi^2)^{1/3}}{32} \frac{n}{\xi} \left\{ (2\pi^2)^{1/3} - \frac{\xi^2}{n} \left[\frac{\hbar c}{E_{\rm sym}(n,T=0)}\right]^3 \right\},
\end{equation}
as in eq.~(\ref{eq:YpBInv}), and where $\xi$ is defined as
\begin{equation}
\xi \equiv \left[ \frac{E_{\rm sym}(n,T=0)}{\hbar c} \right]^2  
\left\{ 24 n \left[ 1+ \sqrt{ 1 +  \frac{\pi^2 n}{288}\left(\frac{\hbar c}{E_{\rm sym}(n,T=0) }\right)^3}\right]  \right\}^{1/3},
\end{equation}
as in eq.~(\ref{eq:xi}).

Thus, if the form of $E_{\rm sym}(n)$ is known, this will uniquely specify $Y_{p,\beta}$. Alternatively, if the $\beta$-equilibrium proton fraction is known from the cold EOS, it can be used to fit for the parameters of the particular model of $E_{\rm sym}$. In the context of this paper, specifying $Y_{p,\beta}$, $S_0$, and $L$ can be used to fit for the parameter $\gamma$; or, specifying $S_0$, $L$, and $\gamma$ can be used to calculate $Y_{p,\beta}$. The latter option is particularly useful as it allows our framework to be applied to parametric EOS that may not calculate $Y_{p,\beta}$ directly.

Finally, we provide the derivative of $Y_{p,\beta}$, which is required to calculate the sound speed in Appendix B. The derivative at constant entropy is given by
\begin{equation}
\frac{\partial Y_{p,\beta}}{\partial n}\biggr\rvert_S = \frac{1}{16} \left(\frac{\pi}{2}\right)^{2/3} 
\left\{ (2\pi^2)^{1/3}\left( \frac{1 - n \phi}{\xi} \right) + \xi \left[ \frac{\hbar c}{E_{\rm sym}(n,T=0)} \right]^3 \left( 3x -\phi \right) \right\}
\end{equation}
where for simplicity we have introduced the quantities
\begin{equation}
\phi = \frac{1}{2 n} \left( 3 x n + 1 \right) +
 \frac{1}{6n} \left\{ 1+\frac{\pi^2 n}{288} \left[ \frac{\hbar c}{E_{\rm sym}(n,T=0)} \right]^3 \right\}^{-1/2} (3xn - 1)
\end{equation}
and
\begin{equation}
x = \frac{1}{n^2} \frac{P_{\rm sym}(n,T=0)}{E_{\rm sym}(n,T=0)}.
\end{equation}

\section{B. Calculation of the Sound Speed}
\label{sec:causality}


In this paper, we have provided the complete set of expressions necessary to extend any cold EOS to non-equilibrium conditions and arbitrary temperature. These expressions can be used to create a new finite-temperature EOS, by varying either the cold, underlying EOS or any of the five parameters of our model. In creating a new EOS, it is useful to always to check that the choice of parameters results in a model that is causal at all densities and temperatures of interest. To that end, we here provide a sample calculation of the adiabatic sound speed for our model.

The sound speed will need to be calculated differently depending on the relevant timescales for the astrophysical system at hand. If the sound-crossing timescale is longer than the time for weak interactions, then matter will remain in $\beta$-equilibrium as the system evolves and the proton fraction will change accordingly. This scenario may correspond to the early phases of a neutron star merger or the cooling of proto-neutron stars. Alternatively, if the dynamical timescale is shorter than the timescale required to maintain $\beta$-equilibrium, as in the late stages of a merger, the proton fraction will remain approximately constant. 

In this appendix, we will calculate the sound speed for the latter case: of a system with a constant proton fraction. For such a system, the adiabatic sound speed, $c_s$, is defined as
\begin{equation}
\left( \frac{c_s}{c}\right)^2 \equiv \frac{\partial P(n, T)}{\partial \epsilon} \biggr\rvert_{S,Y_p}
\end{equation}
where $\epsilon \equiv E(n,T)n + m c^2 n$ is the relativistic energy density, consisting of the classical internal energy density and the rest mass density.  Here, we have suppressed the proton-fraction dependence of the pressure and energy models because we are considering a system that maintains its initial proton fraction, $Y_{p,\beta}(n)$. We can expand this derivative as follows
\begin{equation} 
\left( \frac{c_s}{c}\right)^2  = \frac{\partial P(n, T)}{\partial n} \biggr\rvert_{S,Y_p} \left( \frac{\partial \epsilon}{\partial n} \biggr\rvert_{S,Y_p}\right)^{-1} =  \frac{\partial P(n,T)}{\partial n}  \biggr\rvert_{S,Y_p} \left[ \frac{n}{E(n, T) n + P(n, T) + mc^2n} \right],
\end{equation}
For each term in the pressure expressions of Box II., we calculate and provide the derivatives at constant entropy below.

For a tabular, cold EOS in $\beta$-equilibrium, the cold pressure derivative must be calculated numerically. For a polytropic cold EOS, however, the derivative is simply
\begin{equation}
\frac{\partial P_{\rm cold}(n, T=0)}{\partial n} \biggr\rvert_{S,Y_p} = \frac{\Gamma}{n} P_{\rm cold}(n, T=0),
\end{equation}
where $\Gamma$ is the polytropic index. In the following expression for the complete derivative, we assume the cold EOS can be represented as a polytrope. However, if this is not the case, the first term  should simply be replaced by the numerical derivative of the cold EOS. The total derivative for the case of a constant proton fraction is then

\begin{align}
\begin{split}
\frac{\partial P(n, T)}{\partial n} \biggr|_{S,Y_p} =
    \frac{\Gamma}{n} P_{\rm cold}(n, T=0) 
 + 
\begin{cases}
 16 f_s \sigma T^4/(9c n) , 						 &\phantom{n_1<} n < n_1 \\
 5T/3 								 &n_1 < n < n_2 \\
  \frac{ \partial P_{\rm th, deg}(n,T)}{\partial n} \biggr|_{S,Y_p}  &\phantom{n_1<}  n > n_2 .
\end{cases}
\end{split}
\end{align}

The degenerate thermal pressure of nucleons and electrons, $P_{\rm th, deg}$, is given for symmetric matter by
\begin{equation}
P_{\rm th, deg}(n, T)  = - \left[ a_{\rm SM}\prime + a_e\prime  Y_{p,\beta} \right] n^2 T^2,
\end{equation}
as in Box II. We assume that this is approximately equal to the $\beta$-equilibrium expression because the thermal symmetry-energy correction is small, as discussed in $\S$\ref{sec:thermal}. This assumption will likely introduce a small error into the final sound speed, which we neglect for the present purposes.

In this appendix, for brevity, we will use the following notation: $a_{\rm SM} \equiv a(0.5 n, 0.5 M^*_{\rm SM})$ and $a_{\rm e} \equiv a(Y_{p,\beta} n, m_e)$. Additionally,
\begin{subequations}
\begin{align}
a_{\rm SM}\prime & \equiv \frac{\partial a(0.5 n, 0.5 M^*_{\rm SM})}{\partial n}\biggr\rvert_{Y_q} = 
-\frac{2 a_{\rm SM}}{3 n} 
\left\{ 1 -\frac{B}{2} \left(C+ 3 A \right) \right\} ,
\\
a_{\rm e}\prime & \equiv \frac{\partial a(Y_{p,\beta} n, m_e)}{\partial n}\biggr\rvert_{Y_q} = 
-\frac{2 a_{\rm e}}{3 n} 
\left\{ 1 -\frac{BC}{2} \right\} ,
\end{align}
\end{subequations}
where we have introduced
\begin{subequations}
\begin{align}
A & \equiv  A(n_q, M^*) = \frac{\partial \ln{ M^*(n_q)  }}{\partial \ln{n}} \biggr\rvert_{Y_q} = -  \alpha \left[ 1- \left(\frac{ M^*(n_q)}{mc^2}\right)^2 \right] 
\\
B & \equiv  B(n_q, M^*) =  \frac{M^*(n_q)^2}{M^*(n_q)^2 + (3\pi^2 n_q)^{2/3}(\hbar c)^2}
\\
C & \equiv  C(n_q, M^*) =\frac{(3\pi^2 n_q)^{2/3}(\hbar c)^2}{M^*(n_q)^2}.
\end{align}
\end{subequations}
For the symmetric nuclear terms, $n_q \rightarrow 0.5 n$ and $M^*(n_q) \rightarrow 0.5 M^*_{\rm SM}(0.5 n)$. For the lepton term, $n_q \rightarrow Y_{p,\beta}n$ and the effective mass is simply the electron mass.

Using the entropy expressions of $\S$\ref{sec:boxes}, we can then write the derivative of the degenerate thermal pressure as
\begin{align}
\begin{split}
\frac{\partial P_{\rm th, deg}(n, T)}{\partial n} \biggr|_{S,Y_p} =& -\left[ \frac{\partial (a_{\rm SM}\prime)}{\partial n}\biggr\rvert_{S,Y_p} + \frac{\partial (a_{\rm e}\prime) }{\partial n}\biggr\rvert_{S,Y_p} Y_{p,\beta}  \right] n^2T^2 
+ 2 P_{\rm th, deg}(n,T) \left[ \frac{1}{n} - \frac{ a_{\rm SM}\prime  + a_{\rm e}\prime \cdot Y_{p,\beta} }{a_{\rm SM} +a_{\rm e} Y_{p,\beta}} \right]
\end{split}
\end{align}

The second derivative of $a_{\rm SM}$ is given by


\begin{equation}
\label{eq:ddaNucl}
\frac{\partial (a_{\rm SM}\prime) }{\partial n}\biggr\rvert_{S,Y_p} = a_{\rm SM}\prime \left( \frac{a_{\rm SM}\prime}{a_{\rm SM}} - \frac{1}{n} \right) + \frac{2a_{\rm SM}}{3n^2} B \left[ 3 A^2  - \frac{1}{3} B(3A+C)^2 + \frac{1}{3} C + \frac{3n}{2} \frac{\partial A}{\partial n} \right]
\end{equation}
The second derivative of the electron term, for which $A\rightarrow 0$ due to its constant effective mass, is simply
\begin{equation}
\label{eq:ddaL}
\frac{\partial (a_{\rm e}\prime)}{\partial n}\biggr\rvert_{S,Y_p} = a_{\rm e}\prime 
\left( \frac{a_{\rm e}\prime}{a_{\rm e}} -\frac{1}{n} \right)
+ \frac{2a_{\rm e}}{9n^2} 
BC(1-BC)
\end{equation}
Finally, the second-derivative of the $M^*$ term is given by
\begin{equation}
\frac{\partial A}{\partial n} 
	= \frac{\partial}{\partial n}\left[ \frac{\partial \ln M^*(n_q)}{\partial n} \right] 
	= \frac{2 \alpha}{n} \left[ \frac{M^*(n_q)}{Y_qmc^2}\right]^2\left[ \frac{\partial \ln M^*(n_q)}{\partial n} \right],
\end{equation}
where we have assumed that $M^*$ is defined here for symmetric matter, so that $M^*(n_q) \rightarrow 0.5 M^*_{\rm SM}(0.5 n)$.

\end{document}